\documentclass[acmsmall]{acmart}

\usepackage{amsmath} 
\usepackage{graphicx} 
\usepackage{subcaption} 
\usepackage{url} 
\usepackage{tabularx} 
\usepackage{multirow}   
\usepackage{booktabs}   
\usepackage[acronym,toc]{glossaries} 
\usepackage{tikz}
\usepackage{footnote}

\acmJournal{TOIT}
\acmVolume{VV}
\acmNumber{NN}
\acmArticle{111}
\acmMonth{1}
\acmYear{2024}
\acmDOI{10.1145/1122445.1122456}

\title{MACH: Multi-Agent Coordination for RSU-centric Handovers}

\author{Nikolaus Spring}
\authornote{This work was conducted during his Master's thesis at TU Wien, prior to his current position at Amazon AWS, Germany.}
\email{nikolaus.spring1@gmail.com} 

\affiliation{%
  \institution{Distributed Systems Group, TU Wien}
  \streetaddress{Argentinierstraße 8}
  \city{Vienna}
  \state{Vienna}
  \country{Austria}
  \postcode{1040}
}

\author{Andrea Morichetta}
\email{a.morichetta@dsg.tuwien.ac.at}
\orcid{0000-0003-3765-3067}
\affiliation{%
  \institution{Distributed Systems Group, TU Wien}
  \streetaddress{Argentinierstraße 8}
  \city{Vienna}
  \state{Vienna}
  \country{Austria}
  \postcode{1040}
}

\author{Boris Sedlak}
\email{b.sedlak@dsg.tuwien.ac.at}
\orcid{0009-0001-2365-8265}
\affiliation{%
  \institution{Distributed Systems Group, TU Wien}
  \streetaddress{Argentinierstraße 8}
  \city{Vienna}
  \state{Vienna}
  \country{Austria}
  \postcode{1040}
}

\author{Schahram Dustdar}
\email{dustdar@dsg.tuwien.ac.at}
\orcid{0000-0001-6872-8821}
\affiliation{%
  \institution{Distributed Systems Group, TU Wien}
  \streetaddress{Argentinierstraße 8}
  \city{Vienna}
  \state{Vienna}
  \country{Austria}
  \postcode{1040}
}

\begin{abstract}
This paper introduces MACH, a novel approach for optimizing task handover in vehicular computing scenarios. To ensure fast and latency-aware placement of tasks, the decision-making -- \textit{where} and \textit{when} should tasks be offloaded -- is carried out decentralized at the Road Side Units (RSUs) who also execute the tasks. By shifting control to the network edge, MACH moves away from the traditional centralized or vehicle-based handover method. Still, it focuses on contextual factors, such as the current RSU load and vehicle trajectories. Thus, MACH improves the overall Quality of Service (QoS) while fairly balancing computational loads between RSUs. To evaluate the effectiveness of our approach, we develop a robust simulation environment composed of real-world traffic data, dynamic network conditions, and different infrastructure capacities. 
For scenarios that demand low latency and high reliability, our experimental results demonstrate how MACH significantly improves the adaptability and efficiency of vehicular computations. By decentralizing control to the network edge, MACH effectively reduces communication overhead and optimizes resource utilization, offering a robust framework for task handover management.

\end{abstract}



\begin{document}

\newacronym{mas}{MAS}{Multi-Agent System}
\newacronym{mec}{MEC}{Multi-Access Edge Computing}
\newacronym{vec}{VEC}{Vehicular Edge Computing}
\newacronym{ioe}{IoE}{Internet of Everything}

\newacronym{ec}{EC}{Edge Computing}
\newacronym{ran}{RAN}{Radio Access Network}
\newacronym{ar}{AR}{Augmented Reality}
\newacronym{vr}{VR}{Virtual Reality}
\newacronym{ad}{AD}{Autonomous Driving}
\newacronym{qos}{QoS}{Quality of Service}
\newacronym{vm}{VM}{Virtual Machine}
\newacronym{ue}{UE}{User Equipment}
\newacronym{adas}{ADAS}{Advanced Driver Assistance Systems}

\newacronym{rsu}{RSU}{Road Side Unit}
\newacronym{obu}{OBU}{On-board Unit}
\newacronym{v2i}{V2I}{Vehicle-to-Infrastructure}
\newacronym{v2v}{V2V}{Vehicle-to-Vehicle}
\newacronym{i2i}{I2I}{Infrastructure-to-Infrastructure}
\newacronym{dsrc}{DSRC}{Dedicated Short-Range Communication}

\newacronym{ml}{ML}{Machine Learning}
\newacronym{dai}{DAI}{Distributed Artificial Intelligence}
\newacronym{uddi}{UDDI}{Universal Description, Discovery, and Integration}
\newacronym{api}{API}{Application Programming Interface}

\newacronym{vho}{VHO}{Vertical Handover}
\newacronym{mn}{MN}{Mobile Node}
\newacronym{ap}{AP}{Access Point}
\newacronym{sho}{SHO}{Soft Handover}
\newacronym{hho}{HHO}{Hard Handover}
\newacronym{mcdm}{MCDM}{Multi-Criteria Decision-Making}
\newacronym{rss}{RSS}{Received Signal Strength}
\newacronym{gra}{GRA}{Grey Relational Analysis}
\newacronym{its}{ITS}{Intelligent Transportation Systems}

\newacronym{drl}{DRL}{Deep Reinforcement Learning}
\newacronym{mdp}{MDP}{Markov Decision Process}
\newacronym{dqn}{DQN}{Deep Q-Network}
\newacronym{v2x}{V2X}{Vehicle-to-Everything}
\newacronym{saw}{SAW}{Simple Additive Weighting}
\newacronym{mdrlha}{MDRLHA}{Multi-agent Deep Reinforcement Learning-based Hungarian Algorithm}
\newacronym{vmec}{VMEC}{Vehicle-Assisted Mobile Edge Computing}
\newacronym{vnf}{VNF}{Virtual Network Function}
\newacronym{ilp}{ILP}{Integer Linear Programming}
\newacronym{tesu}{TESU}{Total Edge Servers Utilization}
\newacronym{tesat}{TESAT}{Total Edge Servers Allocation Time}
\newacronym{ai}{AI}{Artifical Intelligence}
\newacronym{meo}{MEO}{MEC Orchestrator}
\newacronym{vfn}{V-FN}{Vehicle-Based Fog Node}
\newacronym{cmab}{CMAB}{Contextual Multi-Armed Bandit}
\newacronym{sdn}{SDN}{Software-defined Networking}
\newacronym{rl}{RL}{Reinforcement Learning}
\newacronym{iov}{IoV}{Internet of Vehicles}
\newacronym{iot}{IoT}{Internet of Things}

\newacronym{etsi}{ETSI}{European Telecommunications Standards Institute}
\newacronym{cam}{CAM}{Cooperative Awareness Message}
\newacronym{fmf}{FMF}{Follow Me Fog}
\newacronym{rs}{RS}{RoadSide}
\newacronym{flops}{FLOPS}{Floating Point Operations Per Second}
\newacronym{flop}{FLOP}{Floating Point Operations}
\newacronym{sumo}{SUMO}{Simulation of Urban MObility}
\newacronym{ve}{VE}{In-Vehicle}

\newacronym{arhc}{ARHC}{Agent-based RSU Handover Coordination}
\newacronym{gpu}{GPU}{Graphics Processing Unit}
\newacronym{ho}{HO}{Handover}
\newacronym{snr}{SNR}{Signal-to-Noise Ratio}

\newacronym{bn}{BN}{Bayesian Network}
\newacronym{av}{AV}{Autonomous Vehicle}
\newacronym{uav}{UAV}{Unmanned Aerial Vehicle}
\newacronym{milp}{MILP}{Mixed-Integer Linear Programming}
\newacronym{aco}{ACO}{Ant Colony Optimization}

\maketitle

\section{Introduction}
\label{sec:introduction}

Driving automation has been gaining traction in the last decade, with autonomous driving~\cite{meneguette2021vehicular} representing a captivating and remunerative~\cite{deichman2023ADFuture} market. In this direction, we are witnessing a combination of centralized hardware (i.e., in Cloud centers) with highly distributed software for vehicle operation~\cite{sedlak_slo-aware_2024}.
Furthermore, modern vehicles execute various kinds of workload~\cite{huang2019social,ieee2024ADArchietecture}, from infotainment to smart city tasks, piling up to a unprecedented complex combination.
This draws a computing landscape that is tightly, yet dynamically connected, where vehicles, edge nodes, and central high-performance infrastructure cooperate together to execute diverse workloads.

One main aspect that has been under the spotlight in the last years is the different ways that computation and task execution flows. Various models have been studied, such as Vehicle-to-Vehicle (V2V), Vehicle-to-RSU (V2R), RSU-to-RSU (R2R), RSU-to-Everything (R2X), Vehicle-to-Cloud (V2C), or more broadly Vehicle-to-Everything (V2X)~\cite{meneguette2021vehicular, wang2020digital, zhao_deep_2025}.
The second aspect--equally crucial, but under-explored in comparison to the computation flow-is where the decision on ``what to compute and where'' happens. Explicitly differentiating among these locations clarifies the strategic position of a solution. The coordination of offloading decisions can happen in different parts of the computing continuum environment, each with distinct implications. A common solution is to have centralized (cloud-based) coordination~\cite{huang2019social,ieee2024ADArchietecture}, characterized by powerful computational resources but higher latency~\cite{liu2020comparison} and communication overhead. Edge-based (RSU-level)~\cite{Grislin2020, Huang2020,sedlak_slo-aware_2024, Gutierrez2011, Ouarnoughi2022} decentralized coordination, enabling lower latency and global awareness within local scopes.
Vehicle-based decentralized coordination, offering minimal latency but often limited by vehicle-level computation~\cite{Alkaabi2024} and network awareness.

Our solution aims at handling V2R offloading by optimizing R2R handovers, in an \textit{infrastructure-based}, \textit{RSU-centric} perspective. However, this is not a trivial task.
Generally speaking, Edge computing environments often consist of heterogeneous devices with varying capabilities. Therefore, it is challenging to design~\cite{Huang2020} MAS that effectively operate across diverse devices, each with different computational and communication capabilities. Additionally, managing the scale of the system is challenging~\cite{Shah2021, Carlier2020} as the number of vehicles and edge nodes increases.
Moreover, MAS need a constant exchange of information to update each agent's knowledge, which can represent a substantial communication overhead. Keeping this overhead low is however essential~\cite{Gutierrez2011} to maintain the benefits of the decentralized approach, especially in bandwidth-constrained vehicular networks. Furthermore, the dynamic nature of vehicular environments requires robust and adaptive algorithms capable of handling frequent changes~\cite{Xu2019} in network topology, varying traffic conditions, and fluctuating resource availability.
Recently, significant effort has been profused for offering RSU-centric offloading~\cite{xue_multi-agent_2025, li_coor_2023, li_deep_2020, munawar_cooperative_2023, fan_game-based_2023, ning_intelligent_2021, zhao_deep_2025, fan_deep_2024, shi_task_2023}, with different, elaborate solutions. However, most of the proposed approaches rely on computational and time-demanding algorithms, such as Deep-Learning-enhanced Reinforcement Learning, that need to be trained. Furthermore, most of these solutions require a central controller computing -- and eventually distributing -- a solution.
Conversely, we offer a lightweight, fully decentralized, multi-agent solution for R2R handovers.

\begin{figure}[ht!]
    \centering
    \includegraphics[width=0.77\columnwidth]{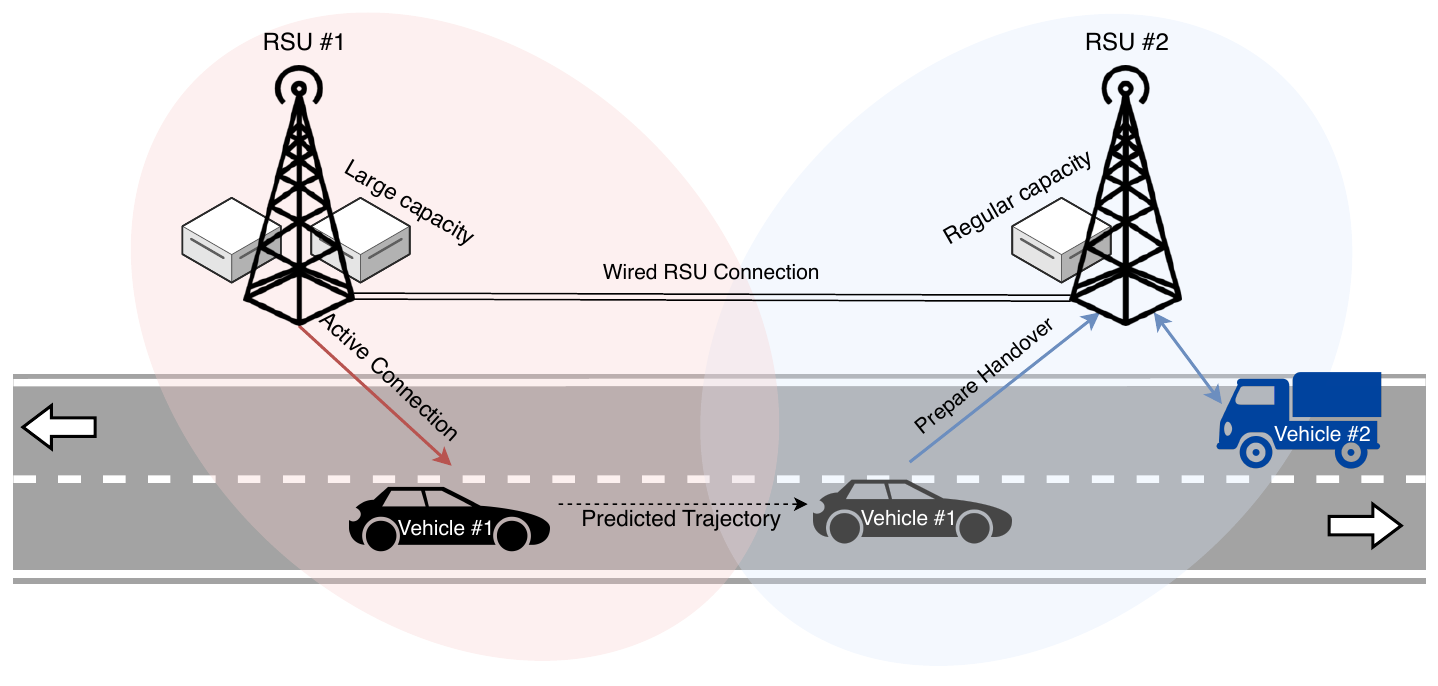}
    \caption{Optimize the Quality of Service (QoS) for vehicular computation offloading: find an RSU station with maximal available capacity and minimum transmission latency; find the best time to handover \textit{Vehicle \#1} from \textit{RSU \#1} to \textit{RSU \#2} according to the predicted vehicle trajectory}
    \label{fig:4_network_layers}
\end{figure}

In particular, this paper presents a robust but lightweight approach for \underline{M}ulti-\underline{A}gent \underline{C}oordination for RSU-centric \underline{H}andovers (MACH). Our \textit{infrastructure-driven} strategy introduces a decentralized, agent-based approach to handover coordination at the RSU layer, moving beyond traditional centralized coordination or vehicle-based methods, and that avoids the overhead of model training and reinforcement loops. The goal is not to compute globally optimal solutions, but to ensure robust, adaptive task handovers that can operate under partial knowledge and real-time constraints in realistic vehicular environments.
We embed agents in the RSUs logic to manage handovers more effectively, optimizing decision-making based on real-time data exchange and vehicle trajectories. Figure~\ref{fig:4_network_layers} gives an overview of how vehicles and RSUs are connected, and how they exchange information.
Furthermore, we provide the agents predictive analytics to minimize unnecessary handovers, reduce communication overhead, and maintain seamless service continuity. This approach ensures that handovers are efficient and occur only when necessary.  dynamically balancing RSUs computational loads using adaptive algorithms, ensuring stable performances across varying traffic conditions and vehicular densities.
We evaluate our strategy against multiple baseline approaches in a simulation environment with real-world traffic datasets. We offer an analysis of how the strategy behaves under different infrastructure and mobility patterns, providing realistic validation of the strategy's effectiveness in managing handovers and optimizing resource allocation.

The remainder of this paper is organized as follows. In Section~\ref{sec:3_relwork} we analyze the state of the art, by exploring related work that integrates MAS into vehicular edge computing. Section~\ref{sec:4_approach} outlines our methodology, explaining the main design decisions. In Section~\ref{sec:setup}, we define the simulation environment scope. We evaluate the robustness of our approach in Section~\ref{sec:6_evaluation} by testing it in three different RSU configurations, and with distinct load-sharing intervals. In Section~\ref{sec:discussion}, we discuss the gist and impact of our proposed solution, underlining future improvements. Finally, in Section~\ref{sec:conclusion} we conclude the paper.

\section{Related Work}
\label{sec:3_relwork}


This section reviews the existing literature on integrating MAS into VEC, focusing on communication strategies for task handover coordination. As a whole, this gives a comprehensive analysis of computation offloading, handover coordination, MAS integration, and security within VEC frameworks. Finally, we outline the research gap left by the presented work and argue how this paper aims to solve it.

\subsection{Computation Offloading in Vehicular Edge Computing}

This section summarizes existing work on computation offloading in VEC, focusing on task-driven optimization techniques and centralized infrastructure-driven solutions. Existing approaches are evaluated based on their ability to ensure strict boundaries in terms of latency, energy consumption, and resource allocation; hence, how apt they are for dynamic vehicular environments.

\subsubsection{Task-driven Offloading Solutions}

\gls{rl} can handle dynamic and uncertain environments where optimal strategies must be learned over time. 
To that extent, Wang et al.~\cite{Wang2020} propose a decentralized offloading algorithm to minimize task completion time in pervasive edge computing. Devices make independent decisions based on partial observations, achieving Nash equilibrium and outperforming other algorithms in efficiency and speed.
Alam et al.~\cite{Alam2022} propose a \gls{mdrlha} for dynamic task offloading in VEC. Their cooperative three-layer VEC network uses moving and parked vehicles as fog nodes, optimizing latency, energy consumption, and cost. 
Lin et al.~\cite{Lin2021} propose a computation offloading strategy for VEC using \gls{drl} to address joint optimization of offloading failure rate and energy consumption. Based on a \gls{mdp}, their model considers task dependencies, vehicle mobility, and diverse computing resources, optimizing using a \gls{dqn}.

Sedlak et al.~\cite{sedlak_slo-aware_2024} provide a collaborative \gls{v2v} offloading mechanism that aims to optimize runtime requirements in vehicle platoons. Using a Bayesian Network (BN) they create a probabilistic view that allows decentralized agents to infer whether a nearby platoon member would be more suitable for running a perception task; in that case, the task is offloaded. 
Wu et al.~\cite{wu_tasks_2022} model the offloading between AVs and RSUs as a multi-objective optimization problem, aiming to reduce processing time and fair load-balancing between RSUs. To solve this problem the apply evolutionary programming.
Kishor and Chakarbarty~\cite{kishor_task_2022} provide a VEC offloading mechanism based on Ant Colony Optimization (ACO) that optimizes latency requirements. However, they assume a static list of devices and fog nodes during operation.
Similarly, Dong et al.~\cite{dong_quantum_2023} provide a multi-task offloading mechanism for Mobile Edge Computing (MEC), optimized through a particle swarm. To explore the solution space they consider each particle's fitness.

Grislin-Le Strugeon et al.~\cite{Grislin2020} embed intelligent agents at vehicles and RSUs for real-time task distribution.
Vehicle agents oversee tasks and assess system load, while RSU agents prioritize requests from less-served vehicle agents. Their simulations show significant improvements in efficiency and fairness, underscoring the importance of agent communication in dynamic VEC networks.
Huang et al.~\cite{Huang2020} introduce an energy-efficient offloading decision-making scheme for VEC that offloads computation-intensive and delay-sensitive applications from vehicles to the network edge. Their dynamic task offloading model minimizes a utility function accounting for energy consumption and packet drop rates; computational resources were allocated based on each vehicle's computation intensity and queue status. Their simulation results show significant reductions in energy consumption and packet drop rates.

\subsubsection{Infrastructure-driven Offloading Solutions}

Alkaabi et al.~\cite{Alkaabi2024} review handover strategies in \gls{mec}, emphasizing the importance of \gls{sdn}-based architectural approaches. They discuss both centralized and distributed architectures for handover management. In centralized systems, a single \gls{sdn} controller handles handovers, leading to potential delays in large networks. 
Ma et al.~\cite{ma_edge_2022} extend vehicular task offloading with an additional layer of UAVs that can ad-hoc improve resource availability in vehicular networks. They optimize their offloading solution through \gls{milp}; through extensive simulations, they showed how UAVs could improve vehicle utility.
Rejiba et al.~\cite{Rejiba2019} propose to assign tasks in VEC using online learning combined with neighbor advice. In their model, tasks are offloaded from vehicles to nearby vehicles with greater computational capacity, managed by an RSUs. Their results show that RSUs using neighbor advice had significantly improved learning performance and reduced task execution delays compared to independent learning.

Ma et al.~\cite{ma_video_2024} present a survey on video offloading in mobile edge computing infrastructures. They documented how smart environments, like smart cities or smart home, have an increasing demand for latency-aware video stream processing; to meet these requirements, they present scenarios in which processing is handed over to one or multiple Edge servers, which can either be static or moving. However, they also point out that the amount on work on this topic is still limited.
Luo et al.~\cite{luo_federated-td3joint_2024} aim to improve the resilience of computing infrastructure by dispatching UAVs that support the overall QoS. In case one of the base stations (i.e., comparable to RSUs) behaves anomalous or fails its service, they extend the infrastructure dynamically with UAVs that hovers over the respective coverage radius.
Xu et al.~\cite{xu_optimization_2023} target idle computing resources and present a collaborative offloading approach where edge devices pay each other to take over their processing task. To avoid resource bottlenecks, they consider historical load distributions to find devices that are likely available.

Hu et al.~\cite{hu_task_2019} present a strategy for optimizing task execution time and energy consumption by offloading computation within an MEC vehicle platoon. Under the assumption of a stable task queue, they present an Lyapunov optimization algorithm that showed to greatly improve the efficacy of task offloading.
Fan et al~\cite{fan_minimum-cost_2019} outline a study on the offloading decision for collaborative task execution between a vehicle platoon and a MEC server, formulating the problem as a shortest path search on a directed acyclic graph. Their approach, using the Lagrangian Relaxation-based Aggregated Cost (LARAC) algorithm, significantly reduces task execution costs while ensuring deadlines are met.
Mousa et al.~\cite{mousa_efficient_2022} propose a UAV-based IoT task offloading strategy for scenarios where traditional edge servers are unavailable, optimizing UAV energy use and task delays through dynamic clustering. They develop a discrete differential evolution (DDE) algorithm for clustering and use ant colony optimization (ACO) to determine the shortest UAV traversal path.
Du et al.~\cite{du_cooperative_2020} present a platoon-based cooperative sensing architecture and a vehicular edge serving scheme to improve real-time traffic data processing for autonomous vehicles. By leveraging the unused computational resources of smart vehicles, their approach efficiently schedules sensing and task offloading, reducing delay costs compared to traditional methods.

\paragraph{RSU-Centric Approaches}

Fan et al.~\cite{fan_deep_2024} focus on a Reinforcement Learning-based, RSU-centric solution. They consider three offloading flow models: V2R, R2R, and R2V. They train and test a TD3-based Deep RL, approach, which guarantees distributed execution for RSU-level decisions. However, it still requires centralized training; plus, the distribution seems to be still seen mostly as a way for execution parallelization rather than a decentralized approach.
Another application of a multi-agent version of TD3 is in the work of Xue et al.~\cite{xue_multi-agent_2025}, where they include the V2V offloading possibility in their analysis. This approach also claim a distributed execution, even though the focus seems to be on parallelization.
Other approaches explore V2R and R2R offloading offering a combination of Deep Learning solutions~\cite{li_deep_2020, li_coor_2023, lu_cooperative_2022, ning_intelligent_2021,lv2021task}. Still, these approaches are mostly centralized.
Differently from the just cited solutions, Fan et al.~\cite{fan_game-based_2023} offer a game-theory-based decision mechanism at the RSU level. However, the paper doesn't discuss where it should be executed and if in a decentralized manner.
In contrast, Zhao et al.\cite{zhao_deep_2025} focus on RSU-to-Vehicle (R2X). Still, they use DRL in combination with a LSTM. They propose a distributed solution where nodes independently select next-hop nodes and task allocations.

\subsection{Simulation Environments for Offloading in VEC}

To analyze parameters such as energy consumption, latency, and resource allocation, simulation tools are essential for evaluating task offloading strategies in VEC  As such, 
\textit{iFogSim}~\cite{Gupta2017} and \textit{CloudSim}~\cite{Buyya2009} are frameworks designed for edge and cloud computing environments. \textit{iFogSim} focuses on optimizing energy consumption and latency, while \textit{CloudSim} models large-scale cloud infrastructures. However, these tools are not tailored for the high mobility context of vehicular networks.
For high mobility scenarios, \textit{Sumo}~\cite{SUMO2018} and \textit{NS-3}~\cite{Riley2010} are more appropriate: \textit{Sumo} enables microscopic traffic simulation, which is critical for implementing and evaluating \gls{v2v} and \gls{v2i} communication approaches. \textit{NS-3} supports simulations of network protocols and communication strategies in dynamic settings, making it ideal for VEC.
Ouarnoughi et al.~\cite{Ouarnoughi2022} develop a comprehensive simulation tool for task offloading in \gls{its} to model vehicle, RSU, and cloud layers. They use a fairness metric to measure the proportion of resources allocated to the least and most served agents, ensuring balanced access to computing resources. However, they do not consider the costs of switching between RSUs, which can impact overall system efficiency. Further, the assumption that a vehicle is always connected to the nearest RSU may not hold true in dynamic and dense urban environments.
Recent studies developed specific simulations for VEC environments to address the unique challenges of \gls{mec}. Their simulation involves vehicles on a straight road using a greedy algorithm for relay node selection, enhancing task processing efficiency: Liu et al.~\cite{Liu2023} utilize vehicular mobility traces from the Europarc roundabout in France to simulate vehicular patterns, providing a realistic context for evaluating task offloading and network performance. Ju et al.~\cite{Ju2023} simulate a complex intersection with vehicles forming \gls{v2v} links, using parameters from the 3GPP standard to model the channel, traffic, and vehicle models. This helps in understanding offloading effectiveness at different processing levels.

\subsection{Takeaways}

Considering the presented work, there exist multiple approaches for offloading computation in vehicular edge computing using multi-agent systems. However, task-driven offloading mechanisms did not adequately address real-time handover complexities or dynamic vehicular mobility patterns. While RL-based approaches, such as those by Wang et al. and Alam et al., improve efficiency, they do not explicitly handle RSU-based handover coordination. Additionally, infrastructure-driven offloading solutions primarily focus on static infrastructures or UAV-supported VEC environments, leaving a gap in effective RSU-level handover optimization. Prior MAS-based strategies, such as those by Grislin-Le Strugeon et al. and Huang et al., demonstrate improvements in task distribution and energy efficiency but lack a comprehensive framework for predictive handover coordination. Furthermore, existing simulation tools, while valuable for assessing offloading efficiency, do not sufficiently model RSU-centric handover processes for real-time vehicular communication.

The research presented in this paper fills this gap by introducing a decentralized, agent-based coordination strategy for RSU-driven handovers. Unlike prior approaches that either rely on vehicle-based or centralized decision-making, ARHC embeds MAS directly in the RSU logic to enable real-time, predictive analytics for optimizing handovers. By incorporating vehicle trajectory prediction and adaptive load balancing algorithms, ARHC minimizes unnecessary handovers and reduces communication overhead while maintaining seamless service continuity. This method ensures robust performance across different vehicular densities and traffic conditions. Moreover, ARHC enhances fault tolerance through distributed agent collaboration, allowing the system to recover dynamically from failures without relying on a single point of control. Its evaluation against baseline approaches using real-world traffic datasets provides a more realistic and thorough validation of its effectiveness in handling complex, dynamic vehicular environments.

~

\section{Methodology}

\label{sec:4_approach}

To optimize QoS in low-latency handover processes, this section introduces MACH -- a lightweight multi-agent coordinator for RSU-based handovers -- which operates decentralized in VEC.
In the following, we describe its underlying networking model, the agent-based communication layer, and the handover decision-making. Thus, we address the limitations of existing methods by enhancing load distribution, reducing unnecessary handovers, and ensuring QoS across dynamic VEC environments.

\subsection{Vehicular Platform and Infrastructure} \label{sec:4_appr:network_model}

The infrastructure for MACH consists of two main layers: the Edge layer, which contains a set of vehicles, and the Fog layer, which contains a set of RSUs as supporting infrastructure. 

\paragraph{Vehicular Edge Layer}

While moving through road networks, vehicles use various services to optimize their operation, e.g., traffic routing or object detection. To use these services with ultra-low latency, vehicles can be extended with Edge devices that allow executing services locally. However, only a limited part can be processed directly on the vehicle, so the remaining load must be handed over to the infrastructure. To ensure low latency, this excluded handing tasks over to the Cloud; hence, the only way to ensure QoS during processing is to find a nearby RSU for handing over the excess load.

\sloppy 
\textbf{Vehicle Definition}: We define a vehicle's state through $s(v) = \langle p,s,d, r \rangle$, where the respective attributes describe the current \textit{position} $p$ of the vehicle, and the \textit{speed} $s$ in which it moves towards a \textit{direction} $d$; finally, $r$ describes the vehicle's integrated computing \textit{resources}.
\fussy

\paragraph{Infrastructure RSU Layer}

RSUs are deployed at fixed locations along the road network, each outfitted with a MEC server. To support the vehicles within their coverage radius, RSUs can perform computations on their behalf. 
The RSUs are interconnected via a wired network, as shown in Figure~\ref{fig:4_network_layers}, which forms an efficient Infrastructure-to-Infrastructure (I2I) communication channel with minimal network congestion. This I2I communication is essential for sharing the load of RSUs and coordinating their actions, which will allow them to make decentralized decisions.

\textbf{RSU Definition}: We define an RSU through $ rsu = \langle p, c ,r \rangle$, which describes the geographical \textit{position} $p$ of the RSU in the road network, its coverage \textit{radius} $r$ for attending vehicles with full QoS, and its maximum communication \textit{capacity} $c$ for receiving handover tasks.

\paragraph{Vehicle-Infrastructure Communication}

Vehicles communicate wirelessly with nearby RSUs, focusing solely on \gls{v2i} communication, represented by the red and blue lines in Figure~\ref{fig:4_network_layers}. This approach reduces complexity, as \gls{v2i} communication scales with the number of vehicles and RSUs, unlike \gls{v2v} communication, which scales quadratically with the number of vehicles.

\paragraph{Limitations}

For simplicity, we do not use established communication models like \gls{rss} or bandwidth. Instead, a fixed range and known positions of RSUs and vehicles are used for calculating their distances. Further, 
this paper does not directly address cellular networks and protocols but builds upon technologies like 5G and WIFI 802.11p. Instead, it abstracts these underlying technologies to focus on effectively coordinating handovers between RSUs.

\subsection{Agent-Based Collaboration} \label{sec:4_appr:agent_architecture}

The presented architecture includes two types of agents -- vehicle agents and RSU agents -- which support decentralized decision-making at vehicles and RSUs.
Vehicle agents are executed on an onboard computing system; they monitor the vehicle state, take handover decisions, and communicate with nearby RSU agents. Meanwhile, RSU agents operate on the MEC servers within RSUs.
Between vehicles and RSUs there exists no ranking, i.e., they are both organized in a flat hierarchy. Note, that we do not always distinguish between a physical vehicle or RSU and its embedded software agent, but summarize them under the vehicle or RSU. Through the context, it should be clear that properties of the vehicle or RSU, such as position and speed, refer to the physical entity, while tasks, such as communication and decision-making are assumed by the agents.



\subsubsection{Tracking Vehicle States}
\label{sec:4_appr:ve_agent}
\glsreset{cam}

\begin{figure}
    \centering
    \includegraphics[width=0.77\linewidth]{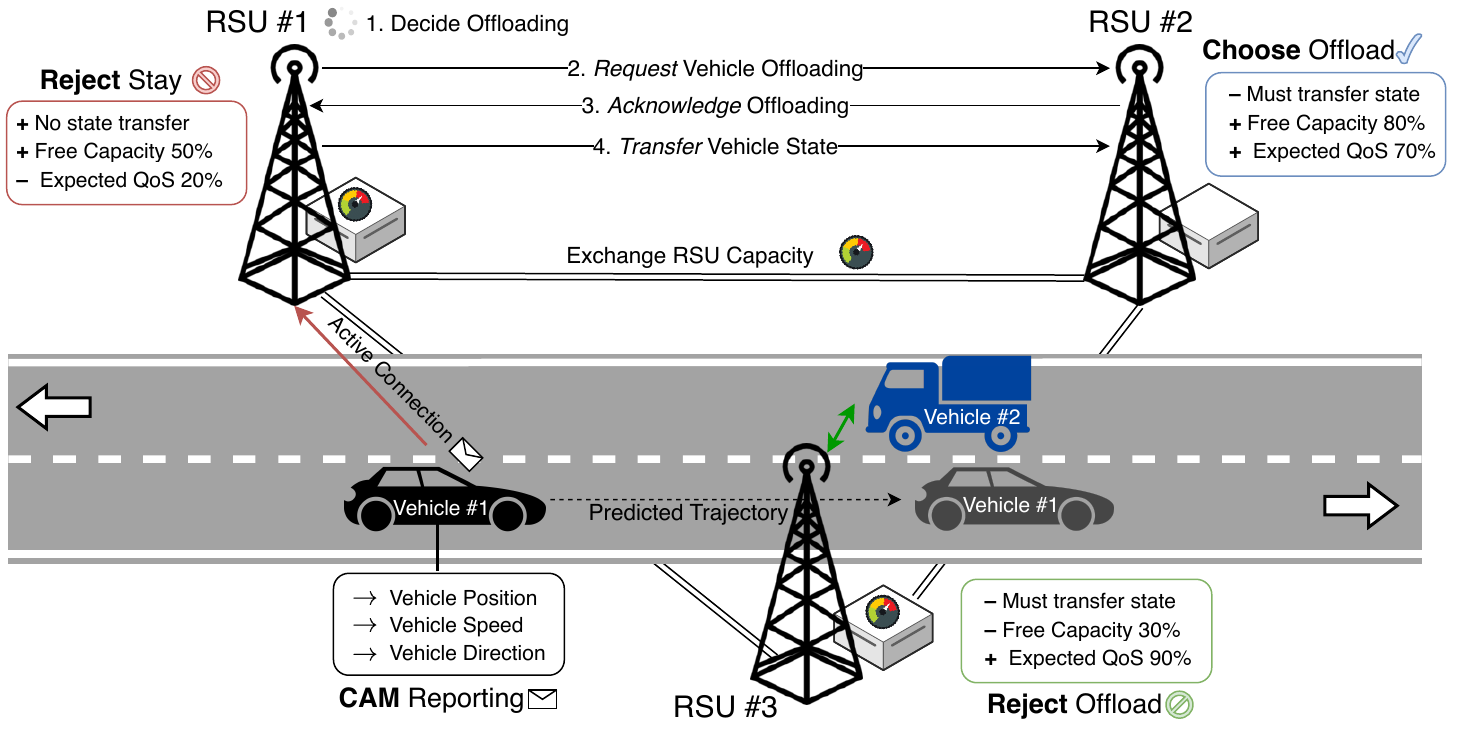}
    \caption{Optimize task handover in vehicular edge computing according to expected QoS and infrastructure load: (1) find best RSU for handover, (2) request handover, (3) accept request, and (4) transfer vehicle state}
    \label{fig:4_appr:rs_agent_communication}
\end{figure}

Vehicles supply their current state to RSUs using \glspl{cam}~\cite{Tesei2021} -- a mechanism developed by the \gls{etsi}. Using \glspl{cam}, vehicles periodically transmit beacon messages to RSUs that contain their position, direction, and speed, i.e., the complete vehicle state $v$. As specified in the \gls{etsi} standard~\cite{etsi:302637-2}, vehicles provide update messages up to 10 times per second.
Thus, RSU agents can track the states of connected vehicles within communication range.
%
This is also depicted in Figure~\ref{fig:4_appr:rs_agent_communication}, which shows how vehicle agents continuously transmit CAMs to their connected RSU agents, allowing RSU agents to monitor the vehicle status in real-time.

\subsubsection{Inter-RSU Communication}

To make informed decisions during handovers, RSU agents periodically exchange their current load and connectivity information with neighboring RSUs. Thus, RSUs are aware if another RSU is about to hit its \textit{capacity} limit. Hence, vehicles and RSUs are constantly updated about the remaining entities. This highlights the importance of the communication rate, in particular, the vehicles' frequency for emitting CAMs, and the frequency of sharing load information among RSU agents. Effective communication management is crucial for balancing network efficiency and reducing congestion, ensuring that necessary information is shared without overwhelming the network. This supports optimal decision-making and coordination among all agents involved in the handover process.

\subsubsection{Vehicular Task Handovers}
\label{subsubsec:vto}

We base handover decisions on two simple but essential aspects: (1) the vehicle trajectory, i.e., which position is expected next, and (2) the load of RSUs -- we summarize this under \textit{connectivity} and \textit{capacity}. If these conditions indicate that another RSU would be more fitting than the one currently serving it, a handover request is sent to the target RSU agent. In the following, we further motivate our choice for these two factors -- connectivity and capacity -- are explain how they are calculated and evaluated during runtime.





\begin{enumerate}
\item \textbf{Connectivity}: To facilitate seamless data transmission, vehicles should remain within the optimal range of RSUs, which is typically influenced by their signal strength and distance. While connectivity is also impacted by other factors, such as bandwidth, network congestion, and variable latencies, these are abstracted in this paper. In particular, we assume that an RSU can provide 100\% QoS within a specified range.
If the vehicle is outside this range, the QoS decreases exponentially as the distance increases; this is expressed in Eq.~\eqref{eq:coverage}:
\begin{equation}
    \label{eq:coverage}
    \text{QoS}_{\text{$v_i$,dist}} = 
    \begin{cases} 
       1 & \text{if } d_i \leq R_j \\
       e^{-\alpha (d_i - R_j)} & \text{if } d_i > R_j
    \end{cases}  
\end{equation}
where $d_i$ is the distance between vehicle $i$ and RSU $j$, $R_j$ is the coverage radius of RSU $j$, and $\alpha$ is a decay constant. For instance, if at 100 meters the QoS drops to 50\% of its maximum value, $\alpha$ can be calculated as: $ \alpha = -\frac{\ln(0.5)}{100-70} \approx 0.0231 $.
\vspace{4pt}

\item \textbf{Capacity Management}: 
Vehicles process part of their load locally, handing over excess load to nearby RSUs, where the aggregated demands from multiple vehicles are handled. Load-based QoS is 100\% if the RSU's load is within its capacity. Note that the RSU's provided QoS directly affects the QoS of all connected vehicles. If the load exceeds the RSU's \textit{capacity}, QoS deteriorates proportionally, calculated as shown in Eq.~\eqref{eq:capacity}:
\begin{equation}
\label{eq:capacity}
    \text{QoS}_{\text{rsu$_j$,load}} = 
    \begin{cases} 
       1 & \text{if } \text{load}_j \leq \text{capacity}_j \\
       \frac{\text{capacity}_j}{\text{load}_j} & \text{if } \text{load}_j > \text{capacity}_j
    \end{cases}
\end{equation}
In a real-world scenarios, it is a common strategy for overburdened RSUs to drop or delay some tasks, leading to a decrease in QoS for all connected vehicles. As resources become strained, the result is lower service quality, resulting in a proportional reduction of QoS.

\vspace{4pt}

\item \textbf{Total QoS}

To combine these two QoS metrics, they are multiplied; hence, a significant drop in either distance-based or load-based QoS severely impacts overall service quality. This combined QoS model captures the impact of both physical distance and the RSU's current load on the vehicle's experience. The combined QoS is calculated as in Eq.~\eqref{eq:total}:
\begin{equation}
\label{eq:total}
    \text{QoS}_{v_i} = \text{QoS}_{\text{$v_i$,dist}} \times \text{QoS}_{\text{rsu$_j$,load}}
\end{equation}

where $\text{rsu}_j$ is the RSU currently connected to vehicle $v_i$.


\end{enumerate}

An example of performing a handover is shown in more details in Figure~\ref{fig:4_appr:rs_agent_communication}: 
\textit{Vehicle \#1} is currently being served by \textit{RSU \#1}, which means that the vehicle sends CAMs to the RSU -- including its position and direction. According to the reported information, \textit{RSU \#1} predicts that \textit{Vehicle \#1} will soon leave its coverage area, which causes the RSU to (1) search for a potential handover candidate. Although \textit{RSU \#3} would have the highest expected QoS, its active connection with \textit{Vehicle \#2} demands a large share of its resources, so that \textit{RSU \#2} ends up the better choice. (2) \textit{RSU \#1} sends a request to \textit{RSU \#2}; (3) upon acceptance, (4) the current state of the vehicle is transferred to \textit{RSU \#2}. This completed the handover from the sending RSU to receiving RSU.


\begin{figure}
    \centering
        \includegraphics[width=0.5\linewidth]{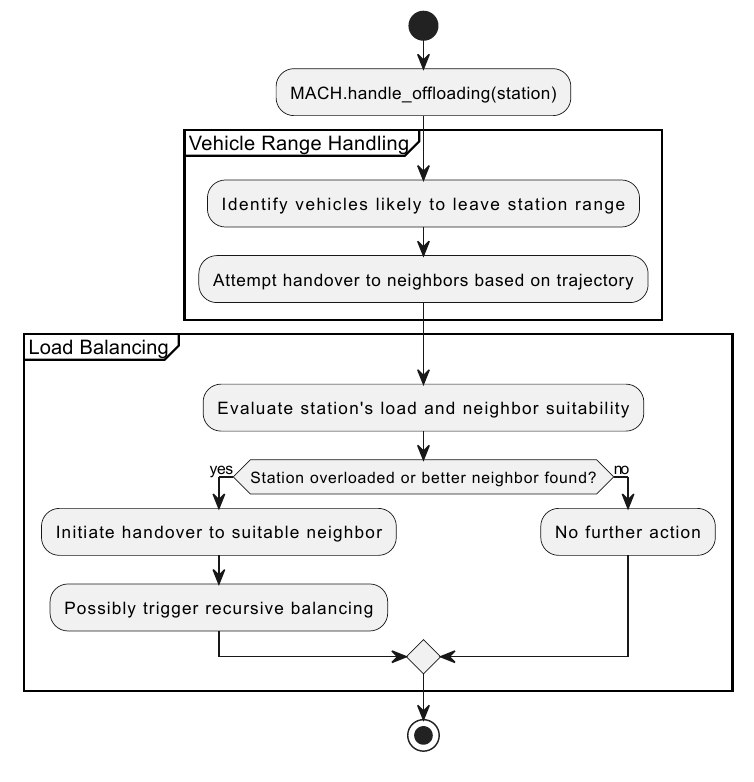}
        \caption{Logic flow of the MACH strategy for vehicles' tasks handovers.}
        \label{fig:uml-activity:MACH-strategy}
\end{figure}

\begin{figure}
    \includegraphics[width=0.85\linewidth]{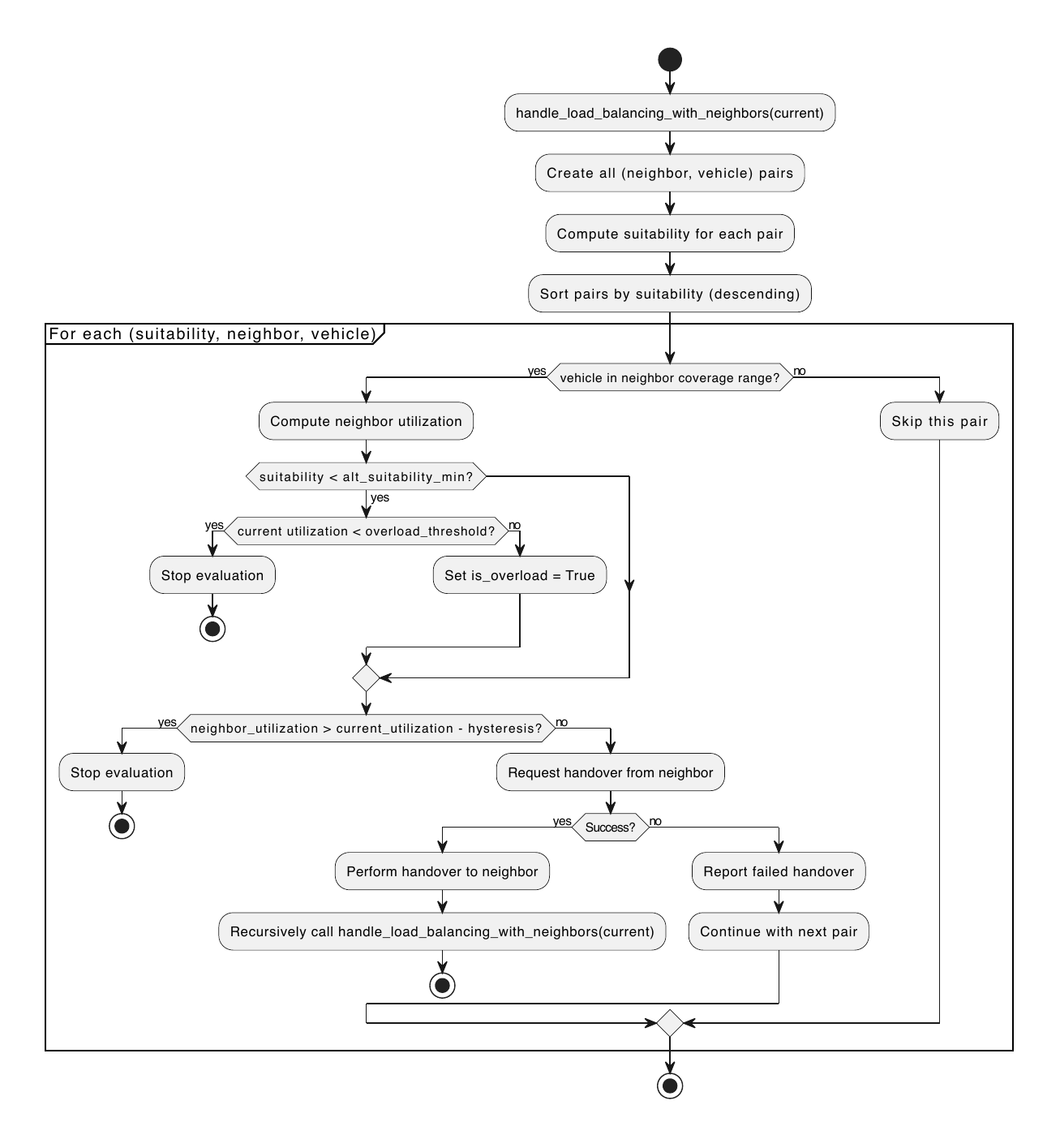}
    \caption{Logic flow of the MACH load balancing strategy.}
    \label{fig:uml-activity:load-balancing}
\end{figure}

More generally, the MACH handover strategy involves migrating the vehicle's computation state from the source RSU to the target RSU without disrupting the service. In the context of this work, we assume a constant state size for all vehicles, focusing on the frequency and effectiveness of handovers rather than the actual data transfer. 
The MACH strategy works in two steps, as depicted in Figure~\ref{fig:uml-activity:MACH-strategy}. First, the RSU agent checks whether there are vehicles that are about to leave its coverage area. Secondly, they control their load and the one of their neighbors, based on the updates that they receive from the other RSU in the VEC architecture. At this point, the agent checks autonomously if there are any more suitable RSUs to handle some of its workloads. The logic is detailed in Figure~\ref{fig:uml-activity:load-balancing}. First, the agent ranks all the neighboring RSUs based on their suitability, which is a function of the vehicle's trajectory and RSU capacity. Based on that, it checks whether a handover would bring more benefits than keeping the task internally.
This simple, but effective strategy aims at having autonomous agents that can take swift and accurate decisions to guarantee the best QoS for the vehicles' tasks.

\section{Experimental Setup}
\label{sec:setup}



To evaluate the presented methodology, we assess the resulting QoS in a dynamic mobility use case. We extend a real-world scenario of a traffic intersection, which allows us to create an accurate simulation environment, where task handover is optimized according to our methodology. Using this environment, we evaluate the impact of MACH on the QoS of moving vehicles, the number of failed/successful handovers, and the inequality of computational load between RSUs.

\subsection{Créteil Roundabout Scenario} \label{sec:6_eval:creteil_roundabout_scenario}

\begin{figure}
    \centering
    \begin{subfigure}[b]{0.30\linewidth}
        \centering
        \includegraphics[width=\linewidth]{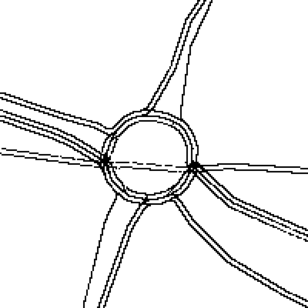}
        \caption{Road topology}
        \label{fig:6_eval:creteil_road_topology}
    \end{subfigure}%
    \hspace{6pt}
    \begin{subfigure}[b]{0.30\linewidth}
        \centering
        \includegraphics[width=\linewidth]{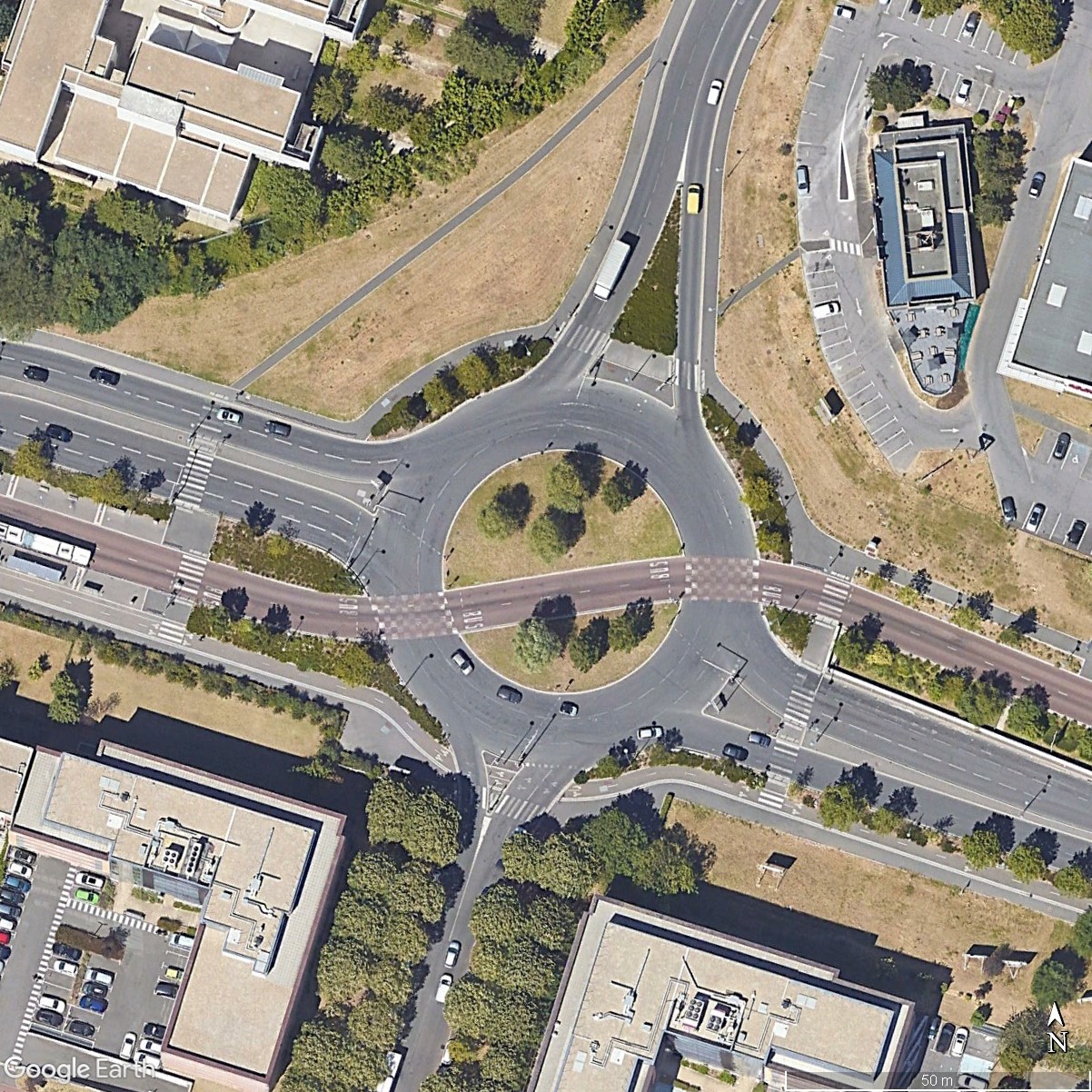}
        \caption{Air view \copyright 2024 Google}
        \label{fig:6_eval:creteil_google_maps}
    \end{subfigure}
    \caption{Europarc Roundabout in Créteil, France}
    \label{fig:6_eval:creteil_topology}
\end{figure}

The \textit{Créteil} Roundabout Dataset~\cite{Lebre2015} provides traffic data for the Europarc roundabout in Créteil, France. The Europarc roundabout is a complex urban intersection with six entry and exit points, with roads consisting of 2 to 3 lanes.
Figures~\ref{fig:6_eval:creteil_topology} illustrates the road topology using both a stylized representation to showcase vehicle positions during the trace analysis, and a real-world reference with an aerial view of the area, sourced from Google Maps. 

Its detailed mobility traces make the Créteil dataset a common choice for analyzing flow and congestion patterns, and evaluating decision-making in mobility scenarios.
For the experiments, the dataset was reduced to an area of 200x200 meters, sufficient to contain the roundabout and significant portions of the entry and exit streets. 
%
%
The Créteil dataset, as available online\footnotemark, contains data captured in the mornings and evenings, i.e., covering peak rush hours. Given that the distributions of the traces are similar, we decided to focus our analysis on the \textit{Morning Dataset}. 

\footnotetext{The Créteil Roundabout Dataset is available at \url{https://vehicular-mobility-trace.github.io/}, accessed on Aug. 3rd, 2024.}

\subsubsection{Morning Dataset}

\begin{figure}
    \centering
    \begin{subfigure}[b]{0.48\linewidth}
        \centering
        \includegraphics[width=\linewidth]{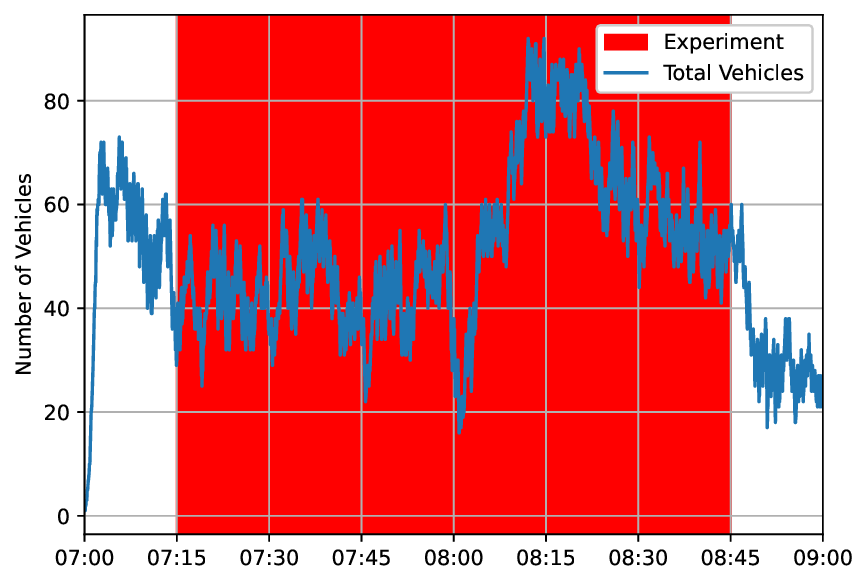}
        \caption{Number of Vehicles Over Time}
        \label{fig:6_eval:creteil_morning_vehicles_over_time}
    \end{subfigure}%
    \hspace{6pt}
    \begin{subfigure}[b]{0.37\linewidth}
        \centering
        \includegraphics[width=\linewidth]{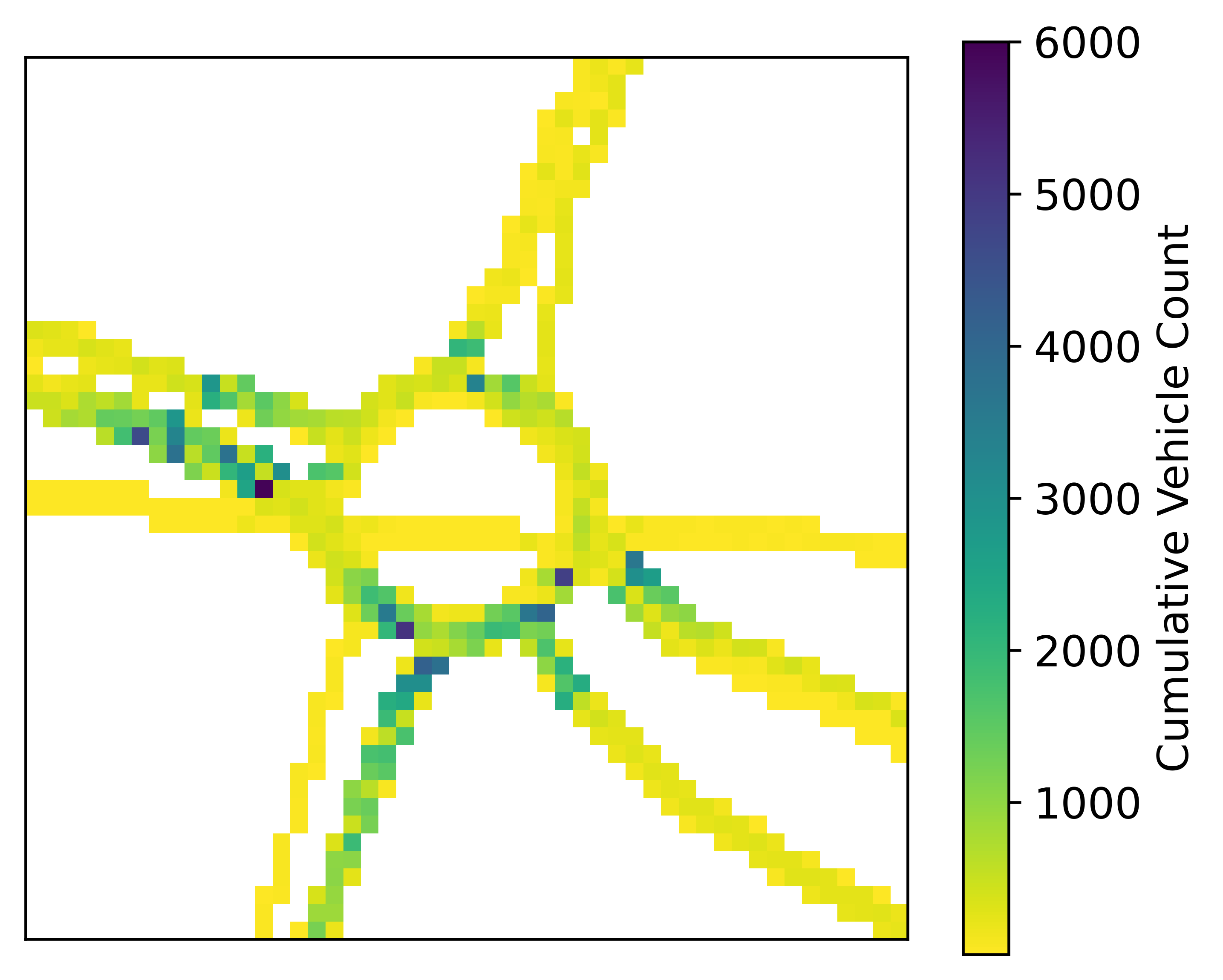}
        \caption{Traffic Density Map}
        \label{fig:6_eval:creteil_morning_density}
    \end{subfigure}
    \caption{Créteil Roundabout, Morning Dataset}
    \label{fig:6_eval:creteil_morning_vehicles_time_density_comb}
\end{figure}

The morning dataset captures traffic data from 7 AM to 9 AM, including approximately 10,000 trips.
Figure~\ref{fig:6_eval:creteil_morning_vehicles_over_time} illustrates the number of vehicles present at different times during the morning dataset period, revealing significant variability in vehicle numbers, with peaks and troughs indicating varying congestion levels.
On average, there were 52 vehicles in the roundabout during the experiment period, with a maximum of 92 vehicles observed at the peak time. Over the entire experiment time frame, 3929 distinct vehicle traces were recorded within the space and time constraints of the roundabout. In our analysis, we consider the time window from 7:15 AM to 8:45 AM (red section in Figure~\ref{fig:6_eval:creteil_morning_vehicles_over_time}) as it reflects peak traffic patterns.
Additionally, Figure~\ref{fig:6_eval:creteil_morning_density} shows the traffic density; for this, we divided the map into 4x4 meter cells, where each cell shows the cumulative vehicle count. Evidently, the hotspots are at the entry roads before the roundabout.


\subsubsection{RSU Placement}

\begin{figure}[h]
    \centering
    \begin{subfigure}[b]{0.30\columnwidth}
        \centering
        \includegraphics[width=\linewidth]{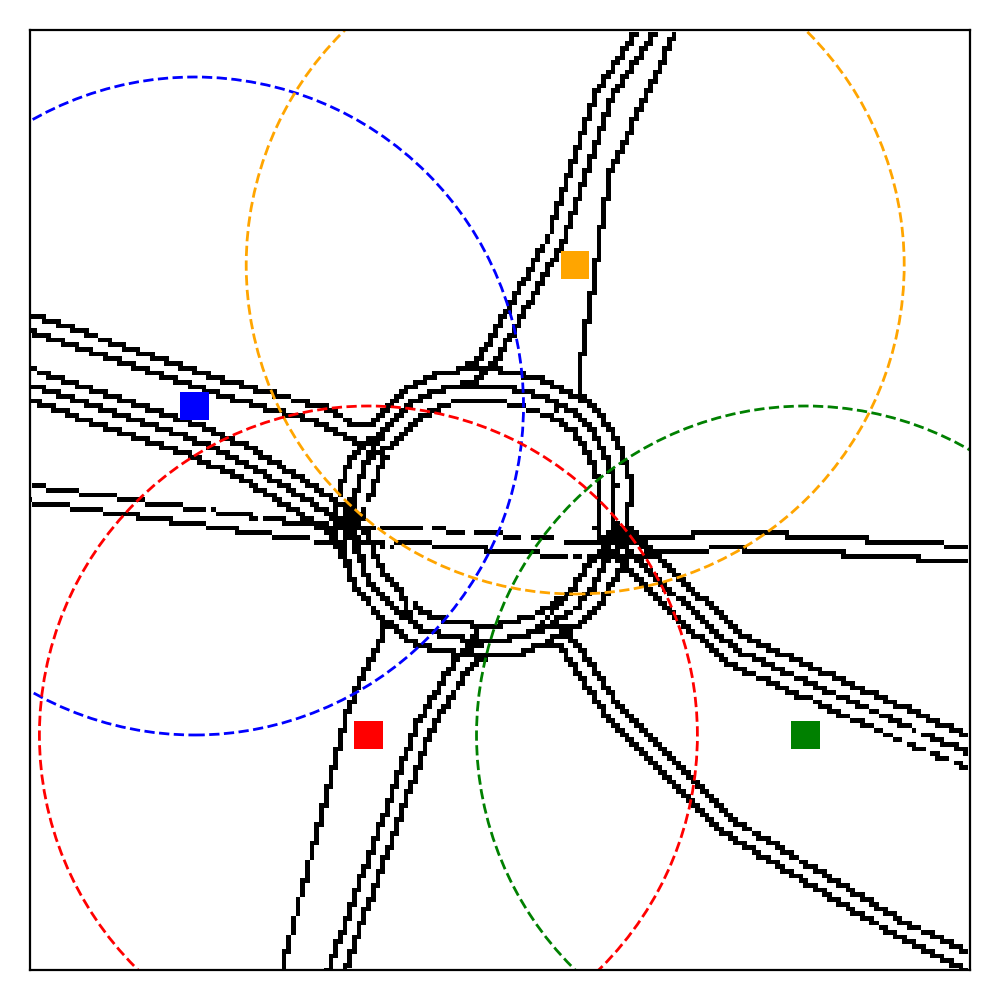}
        \caption{\textbf{Sparse} placement (4 RSUs)}
        \label{fig:6_eval:creteil_4_rsu_config}
    \end{subfigure}
    \hspace{6pt}
    \begin{subfigure}[b]{0.30\columnwidth}
        \centering
        \includegraphics[width=\linewidth]{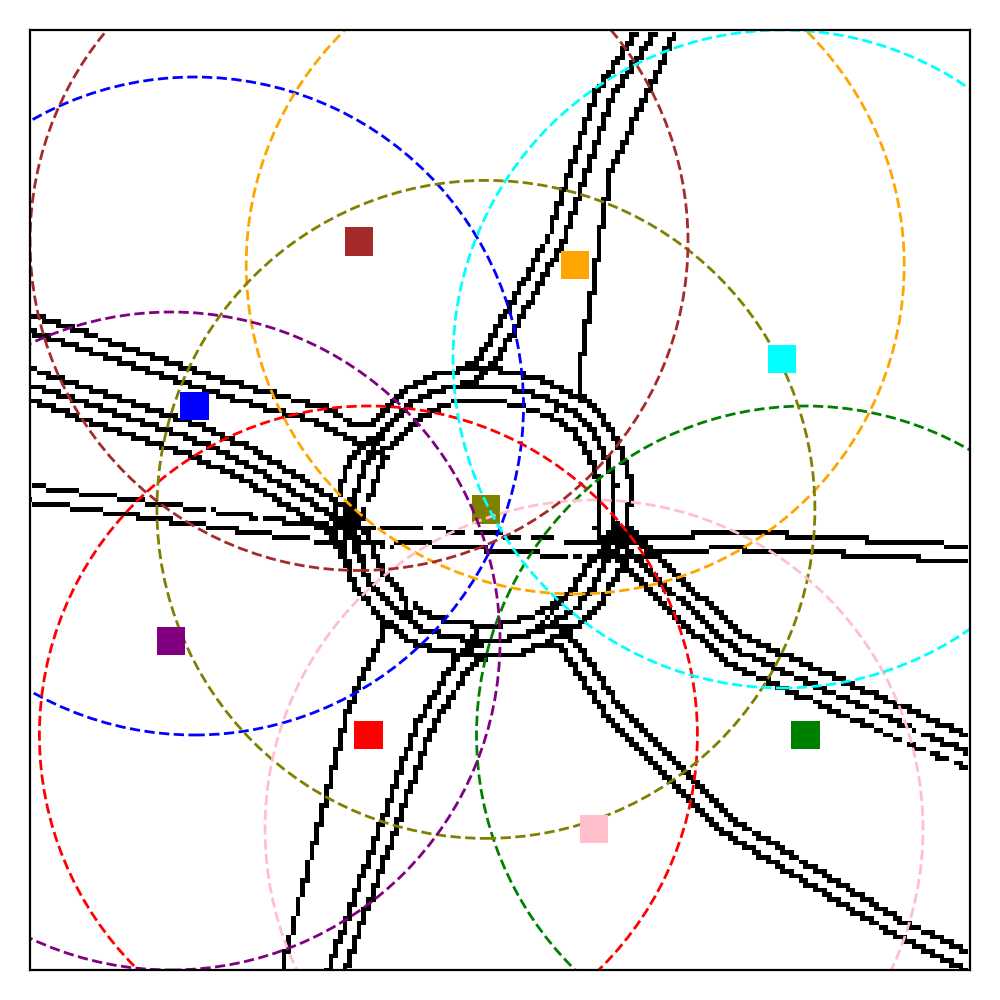}
        \caption{\textbf{Dense} placement (9 RSUs)}
        \label{fig:6_eval:creteil_9_rsu_config}
    \end{subfigure}
    \caption{Comparison of RSU placements for the Créteil Roundabout.}
    \label{fig:6_eval:rsu_configs}
\end{figure}

To evaluate our methodology under varying infrastructure density, we define two RSU placement strategies: a \textbf{sparse} and a \textbf{dense} one, as shown in Figure~\ref{fig:6_eval:rsu_configs}. Additionally, Table~\ref{tab:simulation_parameters} summarizes the key simulation parameters, which provides an overview of the various configurations and settings used throughout the experiments.

\textbf{Sparse RSU placement} \quad  We place four RSUs, each with a range of 70 meters, strategically positioned to cover the roundabout's primary entry and exit points (cfr. Figure~\ref{fig:6_eval:creteil_4_rsu_config}), ensuring a complete coverage of all road segments.
Computations at the RSUs are carried out by Tesla T4 GPUs, each offering 65 tera floating point operations per second (TFLOPS), representing their peak theoretical compute performence. 
To elicit the system boundaries and evaluate its robustness, we add an alternative experimental scenario with \textit{halved} RSU capacity -- 32.5 TFLOPS per RSU. 
This reduced capacity will likely provoke situations where not all vehicles can be provided with 100\% QoS, particularly during high traffic density or localized congestion. Hence, we expect to find deeper insights into the system's performance under limited RSU resources.

\textbf{Dense RSU placement} \quad Nine RSUs, each with a range of 70 meters. This configuration includes the original four RSUs, an additional four positioned between them, and a ninth RSU is placed in the roundabout's center. As shown in Figure~\ref{fig:6_eval:creteil_9_rsu_config}, this arrangement ensures that each road segment is covered by at least two RSUs, with most segments receiving coverage from three or more RSUs.
The \textit{dense} configuration evaluates system performance under a more robust RSU deployment, focusing on enhanced coverage, reduced handover frequency, and improved QoS. By increasing the number of RSUs, we expect to see improvements in load distribution and a more significant reduction of the number of handovers compared to the baseline approaches, compared with the \textit{sparse} setup.
With full capacity, there is a surplus that should maintain 100\% load-based QoS even under peak conditions.
Similarly, we assess the system’s robustness under resource limitations, we add experimental scenarios with \textit{halved} capacity (32.5 TFLOPS) or \textit{quartered} capacity (16.25 TFLOPS).

\textbf{RSU Failure} \quad The scenario uses the sparse RSU setup, with the south RSU deliberately disabled to simulate a failure.
This configuration wants to examine the consequences for the QoS, especially as we chose to disable the RSU from the area that gets more traffic.

\subsection{Simulation Environment}

To develop a simulation environment that closely mirrors real-world VEC scenarios, we designed a custom simulation environment.
While existing simulations, such as ~\cite{Ouarnoughi2022}, model RSU and vehicle agents and implement specific coordination logic, they are tailored to fixed strategies and do not support experimenting with heterogeneous, reconfigurable agent behaviors. To enable a broader investigation of multi-agent interactions under diverse coordination logics, we developed MACH as a modular, extensible agent-based simulation framework.
The core of this simulation is built using the Python Mesa framework~\cite{python-mesa-2020} -- a versatile agent-based modeling and simulation platform. While Mesa provides functionalities similar to well-established MAS simulation frameworks, like NetLogo, Repast, or MASON, one of its advantages is that it simplifies the development by continuously visualizing agent states through an extendable UI. 

The MACH simulation operates with discrete, uniformly spaced time steps. The duration of each step is configurable to align with the sampling rate or resolution of the input dataset. This design segments the vehicles' trajectories time into discrete, fixed-length intervals, treating the system as quasi-static~\cite{Lin2021} within each step while allowing its state to evolve across steps.


\begin{table}
\centering
\caption{Description of the core simulation parameters for the proposed evaluation environment.}
\label{tab:simulation_parameters}
\begin{tabularx}{\columnwidth}{|l|X|} 
\hline
\textbf{Parameter} & \textbf{Value / Description} \\ \hline
\textbf{Scenario} & Créteil Roundabout, Morning Traces \\ \hline
\textbf{Traffic Data Timeframe} & Morning: 7:15 AM - 8:45 AM \\ \hline
\textbf{Simulation Area} & 200x200 meters \\ \hline
\textbf{Number of Vehicles} & Morning: 3929 traces, max 92 concurrent vehicles \\ \hline
\textbf{Vehicle Capacity} & Full: 1.3 TFLOPS\newline Free 70\%: 0.9 TFLOPS \\ \hline
\textbf{RSU Placement} & Sparse: 4 RSUs \newline Dense: 9 RSUs \\ \hline
\textbf{RSU Capacity} & Full: 65 TFLOPS \newline Half: 32.5 TFLOPS \newline Quarter: 16.25 TFLOPS \\ \hline
\textbf{RSU Range} & 70 meters \\ \hline
\textbf{Task Computation Load} & 79.72 GFLOP per frame (2.391 TFLOPS at 30 FPS) \\ \hline
\textbf{Dynamic Load Distribution} & Uniform distribution between 1.9 and 3 TFLOPS, \newline avg. 2.45 TFLOPS \\ \hline
\textbf{CAM Update Frequency} & 1 second (for MACH strategy tuning) \\ \hline
\textbf{RSU Agent Parameters} & Leaving Threshold: 0 \newline Overload Threshold: 0.7 \newline Load-Balancing Hysteresis: 0.05 \newline Minimum Suitability: 0.3 \newline Load information exchange: every 1s, 5s, 10s\\ \hline
\end{tabularx}
\end{table}

\subsubsection{Scenario Development} \label{sec:4_appr:simulation_scenario_dev}

The scenario development for this paper centers on routine operations in urban environments where the density and coverage of RSUs with integrated computing capacity are sufficient to meet the QoS requirements for \gls{ad} tasks or are slightly below the required capacity (in some locations) to test overload behavior. These scenarios balance complexity with manageability by focusing on small areas with intricate street layouts and high-traffic conditions involving numerous vehicles moving simultaneously.

These scenarios are engineered to capture high-fidelity data, recording vehicle positions at least every few seconds. This frequent and precise data collection allows for additional metrics such as speed and orientation, which can also be derived from the detailed location information.


The simulation simplifies the environment by focusing on evaluating handover strategies. It assumes fixed RSUs locations with stable capacities and reliable connections, ideal communication conditions, and static offloading of computational tasks. While primarily focused on normal operations, it also explores RSU failures to assess system robustness.




\subsubsection{Task Computation Model}
\label{sec:6_eval:load_capa_model}

The computational tasks and their respective loads are derived from the work of Ouarnoughi et al.~\cite{Ouarnoughi2022}, who provide a detailed breakdown of the computational requirements for typical autonomous driving tasks. To enhance real-world applicability, we focus on computational loads associated with \gls{ad} tasks, which are critical in VEC environments. Detailed task descriptions are available in the implementation of the task entity module, which can be accessed online\footnote{The task entity implementation is available at \href{https://github.com/Houarnoughi/Simpy-AD/blob/develop/simulation/entity/task.py}{GitHub}, last accessed on August 3, 2024.}.

The complete list of tasks executed by vehicles is summarized in Table~\ref{tab:6_eval:load_ad_tasks}. Since many of these tasks are based on video images, we assume that vehicles operate on their data with 30 FPS, i.e., running 30 computations per second. This frequency is consistent with the studies by Ouarnoughi et al.~\cite{Ouarnoughi2022} and Le-Strugeon et al.~\cite{Grislin2020}, which highlights the importance of high computational throughput for \gls{ad} applications. The total computational load per vehicle at a frame rate of 30 FPS is approximately 79.72 GFLOP per frame, equating to a total of 2.391 TFLOPS per second.

\begin{table}
\centering
\caption{Computational Load for \gls{ad} Tasks}
\label{tab:6_eval:load_ad_tasks}
\begin{tabular}{|l|r|}
\hline
\textbf{Task} & \textbf{Load (GFLOP)} \\
\hline
Traffic Sign Detection & 23 \\
Lane Detection & 0.23 \\
Object Detection & 0.23 \\
Object Tracking & 0.23 \\
Mapping & 0.23 \\
Localization Algorithm & 0.23 \\
Motion Prediction & 0.23 \\
Trajectory Planning & 2.3 \\
Behavior Planning & 2 \\
Route Planning & 50 \\
Control Algorithm & 1 \\
Traffic Light Detection & 0.04 \\
\hline
\textbf{Total Load per Frame} & \textbf{79.72} \\
\hline
\end{tabular}
\end{table}

\subsubsection{Vehicle Onboard Capacity}
Each vehicle is equipped with an NVIDIA TX2 processor, capable of delivering up to 1.3 TFLOPS\footnote{For more information on the device specification, refer to NVIDIA’s official documentation on Jetson modules at \url{https://developer.nvidia.com/embedded/jetson-modules}, last accessed on December 17th, 2024.}. In our model, we assume that each on-board unit can handle up to 70\% of this capacity locally (0.91 TFLOPS), while the remaining load must be offloaded to the RSUs. This approach aligns with standard practices in VEC research, as noted by Ouarnoughi et al. and Grislin-Le Strugeon et al.~\cite{Grislin2020, Ouarnoughi2022}. Although they suggest a more dynamic interaction where vehicles and RSUs communicate to decide which task to offload, we simplify this by assuming that any task exceeding 70\% of the vehicle's capacity is offloaded. 

\subsubsection{Capacity and Connectivity of RSUs}
The RSUs are equipped with NVIDIA Tesla T4 GPUs, providing a peak performance of 65 TFLOPS~\footnote{Further details on the NVIDIA Tesla T4 GPU can be found at \url{https://www.nvidia.com/en-us/data-center/tesla-t4/}, last accessed on August 3, 2024.}. This capacity is chosen to handle the aggregated computational loads from multiple vehicles efficiently. The GPU capacity can be scaled for experimental scenarios requiring varied computational capacities.
%
%
However, for the relatively small scale of the proposed simulation scenarios, we limit the RSU range in our simulations to $d_0 = 70$ meters to ensure practical applicability and model simplicity. The decision is based on empirical data and typical RSU specifications, which suggest that shorter ranges are more realistic and practical for urban scenarios with numerous obstacles~\cite{Lin2021, Liu2021, Liu2023, Rejiba2019}.


\subsubsection{Dynamic Load Distribution}

To introduce more dynamism into the simulation, the load generated by a vehicle is distributed over a uniform distribution of 1.9 to 3 TFLOPS, approximating a $\pm 25\%$ variation. This reflects realistic scenarios where computational demands fluctuate due to varying environmental conditions and task complexities. Each vehicle produces an average of 1.54 TFLOPS that are offloaded to RSUs. A seeded random generator ensures deterministic and reproducible results.

\subsection{Evaluation \& Comparison}

Given the scenario and the simulation environment, we have the main ingredients for running our experimental evaluation. To evaluate the quality of our approach -- the MACH algorithm -- we now present the performance metrics that we capture for each experiment, and afterward, the baseline algorithms that we use to position our results. In that regard, both MACH and the baseline algorithms will be evaluated and compared in terms of the presented performance metrics.

\subsubsection{Performance Metrics}

To evaluate the quality of our approach and the baseline algorithms, we choose four performance metrics: (1) the Gini coefficient -- a measure of how unequal load is distributed among RSUs; (2) the QoS of vehicles -- indicating how satisfying the service is they receive; (3) the number of successful handovers; and (4) the number of failed handovers -- indicating wong offloading decisions. Combined, these four factors reflect the quality of the experiments; in the following, we elaborate them in more detail.

\paragraph{Gini Coefficient}

The equality of load distribution across RSUs is measured using the Gini coefficient, a widely used metric for assessing inequality. A Gini coefficient of 0 indicates perfect equality, while a value of 1 indicates maximum inequality. It is mathematically expressed as:
\begin{equation}
    G = \frac{\sum_{i=1}^{n} \sum_{j=1}^{n} |\text{load}_i - \text{load}_j|}{2n^2 \mu}
\end{equation}
where $\gamma$ is the Gini coefficient, $n$ is the number of RSUs, \(\text{load}_i\) and \(\text{load}_j\) are the loads of the RSU $i$ and $j$ respectively, and $\mu$ is the mean load. In our simulation, we are not capping the load at each RSU's maximum capacity. If, for example, an RSU is supposed to handle 100 but gets assigned 130, we account for that when calculating inequality through the Gini index. Through this modeling choice, we aim to capture the real impact of coordination failures or poor load balancing, as overloads directly increase the inequality measure.

By including excess load in the Gini coefficient calculation, we capture both the distribution of load and the strain on the network when certain RSUs are overloaded. This approach provides a more comprehensive measure of inequality, reflecting not only the fairness of load distribution but also the inefficiencies that arise under stress.
The Gini coefficient effectively evaluates load distribution strategies in vehicular networks. Ensuring even load distribution across RSUs optimizes resource utilization, prevents bottlenecks, and enhances overall network performance. By accounting for load beyond capacity, we gain deeper insights into the network’s behavior under stress, particularly in scenarios where load inequality may lead to RSU overload and QoS degradation.

\paragraph{Quality of Service}

To ensure vehicle tasks are computed with low latency, e.g., for detecting objects, it is crucial to maintain high QoS for all vehicle operations. In Eq~\eqref{eq:total}, we presented how the composite metric $\text{QoS}_{v_i}$ is calculated by combining the capacity and connectivity of RSU connections. In the evaluation, we use both the average and minimum $\text{QoS}_{v_i}$ for highlighting particular strengths and weaknesses of the different algorithms. Overall, this metrics ensures optimal network performance while avoiding service degradation for individual vehicles.

\paragraph{Successful Handovers}

The number of successful handover attempts is a critical metric that counts the instances where RSU agents successfully pass a vehicle task to another RSU agent. A handover is successful when the target RSU agent accepts the request. This metric is crucial for understanding the effectiveness of the handover strategy and the communication overhead in state migration.


\paragraph{Failed Handovers}

This metric tracks instances where a handover attempt is initiated but declined by the target RSU agent. Many failed handovers indicate inefficiencies in the handover decision-making process and contribute to unnecessary communication overhead and lower QoS for the vehicles. This metric helps identify areas where the handover strategy can be optimized.

\subsubsection{Baselines}


To assess the performance of MACH -- our proposed handover strategy -- we compare it against three different benchmark algorithms. We chose the following approaches due to their relevance in related work, but also due to their effectiveness and simplicity.

\begin{enumerate}

\item \textbf{Nearest Handover} \\
For offloading computation, connect a vehicle client always to the nearest RSU; this strategy is also applied by Grislin-Le Strugeon et al.~\cite{Grislin2020} and Ouarnoughi et al.~\cite{Ouarnoughi2022}. Hence, it is a viable way to achieve high QoS by allocating computation to RSUs with low distance.

\item \textbf{Earliest Possible Handover} \\
Offload a task at the earliest possible moment based on the vehicle's distance and the RSU's range. This serves as control benchmarks to measure the performance of more advanced strategies and provide insights into the advantages of timely handover decisions.

\item \textbf{Latest Possible Handover} \\
Similar as above, but offloading vehicle tasks at the latest possible moment. When proactively offloading computation, this strategy takes decision just before vehicles leave their source RSU's coverage area.

\end{enumerate}

Combined, we provided a realistic evaluation scenario which is supported in our custom simulation environment; using this, our approach is compared in multiple experiments against the presented baselines according to performance metrics. In what follows, the results are presented.

\section{Results}

\label{sec:6_evaluation}

In this section, we evaluate the behavior of the proposed architecture with different configurations in comparison with the chosen baseline approaches, i.e., earliest, latest, and nearest-RSU handovers. The analysis focuses exclusively on the morning traffic dataset of the Créteil roundabout, as it exhibits a higher average and maximum number of concurrent vehicles than the evening trace. The evaluation explores two core aspects for our approach -- MACH -- and the baselines: (i) what is the overall number of handovers, (ii) how fair is the load distributed among RSUs, and (iii) what is the QoS perceived by vehicles. We test each strategies in three different scenarios: (a) sparse, with four RSUs, (b) dense, with nine RSUs, and (c) failure, with three RSUs (one out of four fails).

\subsection{Scenario 1: Créteil Roundabout - Sparse RSU Placement} \label{sec:exp:creteil_sparse}

In the first evaluation scenario, we examine the performance of the \textit{sparse} RSU configuration during runtime. Contrarily to the \textit{dense} configuration, fewer infrastructure is needed for operating this scenario, i.e., lower cost. However, this sparse setting might also uncover bottlenecks in the resulting service quality. In the following, we present the results for each evaluated aspect.

\subsubsection{Handover Frequency}

\begin{figure}
    \centering
    \begin{subfigure}[b]{0.45\textwidth}
        \centering
        \includegraphics[width=\linewidth]{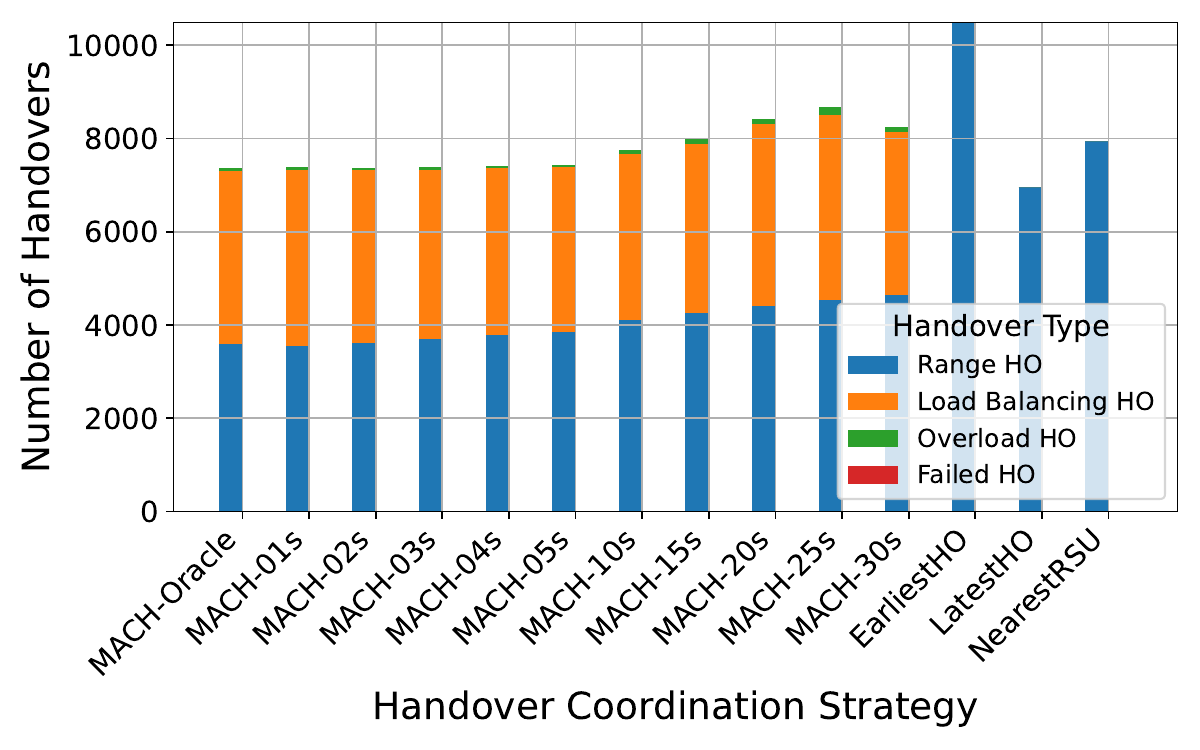}
        \caption{Full RSU Capacity}
        \label{fig:6_eval:res:creteil_morning_sparse_full_ho}
    \end{subfigure}
    \hspace{5pt}
    \begin{subfigure}[b]{0.45\textwidth}
        \centering
        \includegraphics[width=\linewidth]{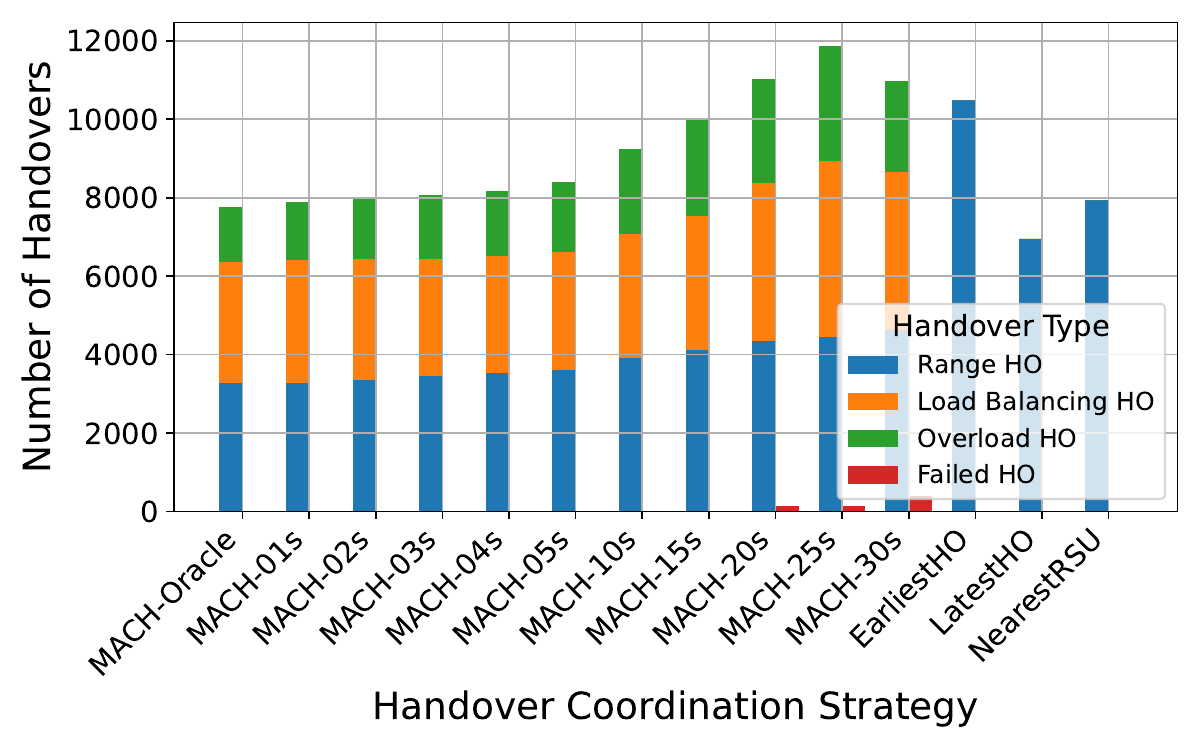}
        \caption{Half RSU Capacity}
        \label{fig:6_eval:res:creteil_morning_sparse_half_ho}
    \end{subfigure}
    \caption{Number of Handovers of Créteil Sparse Configuration at Morning}
    \label{fig:6_eval:res:creteil_morning_sparse_ho}
\end{figure}

Figure~\ref{fig:6_eval:res:creteil_morning_sparse_ho} presents the number of handovers for full and half RSU capacity configurations during the morning mobility traces. Baseline strategies (earliest/latest handover and nearest RSU) are compared alongside MACH, with load-sharing intervals increasing from left to right.
Handovers are classified as successful or failed, with successful handovers further categorized by their triggers: range-based, load-balancing, and overload-triggered.

In the scenario at full RSU capacity (Figure~\ref{fig:6_eval:res:creteil_morning_sparse_full_ho}), the \textit{latest} handover strategy results in the fewest handovers, while the nearest RSU strategy performs slightly worse than MACH with low load-sharing intervals (Oracle util 10 seconds interval); the rationale is presumably that the nearest strategy does not consider other aspects as load balancing, which brings the former strategy to be more conservative. The \textit{earliest} handover strategy generates the highest number of handovers. As the load-sharing interval increases, the total number of handovers rises, driven primarily by more range-based handovers. Importantly, no failed handovers and almost no overload-triggered occur, indicating that the high RSU capacity effectively prevents overloads and ensures reliable handovers.

The baseline strategies maintain a consistent number of handovers regardless of the half/full RSU capacity (Figure~\ref{fig:6_eval:res:creteil_morning_sparse_half_ho}). In contrast, when switching from full to half capacity, MACH results in an increased number of handovers, which is partly driven by overload-triggered handovers. The \textit{latest} handover strategy continues to result in fewer handovers than MACH, while the \textit{earliest} handover strategy sometimes surpasses certain MACH configurations. As the load-sharing interval rises, the total number of handovers increases equally, with a few failed handovers occurring.
Overall, the \textit{latest} handover approach is the one that provides better results in terms of handovers. Clearly, in this scenario, handing over a workload just before the vehicle is about to move to another RSU covering area minimizes the total handover number.
Still, it is essential to understand that this is only one metric in the handover strategy performance evaluation. In the following, we depict other parameters that show a clearer picture. 

\subsubsection{Performance Metrics: QoS and Load Balancing Efficiency}

In this section, we analyze the average QoS and Gini index of MACH and the baseline strategies across various load-sharing intervals.
Furthermore, we explore how the QoS and load balancing efficiency metrics change over time, over varying traffic conditions.

\begin{figure}[h]
    \begin{subfigure}[b]{0.43\linewidth}
        \centering
        \includegraphics[width=\linewidth]{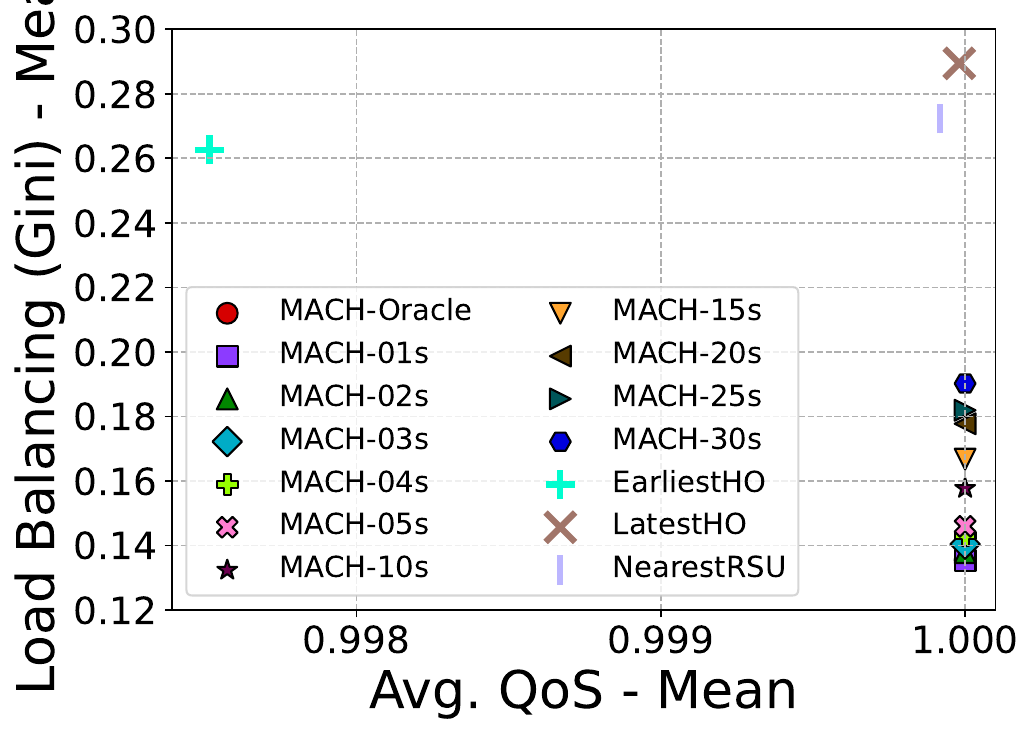}
        \caption{Full RSU Capacity.}
        \label{fig:scatter-sparse-full-morning-avg_gini_qos}
    \end{subfigure} 
    \hspace{5pt}
    \begin{subfigure}[b]{0.43\linewidth}
        \centering
        \includegraphics[width=\linewidth]{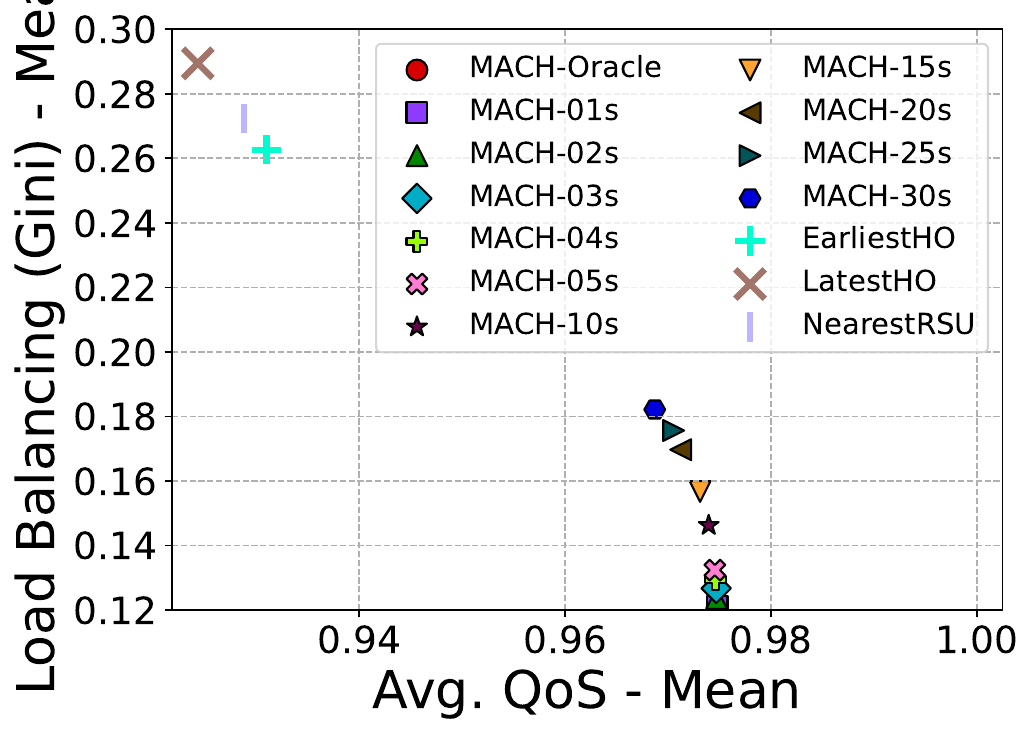}
        \caption{Half RSU Capacity.}
        \label{fig:scatter-sparse-half-morning-avg_gini_qos}
    \end{subfigure}
    \caption{Scatter plot of load balancing and QoS for the sparse RSU scenario.}
    \label{fig:scatter-sparse-avg_gini_qos}
\end{figure}

\paragraph{Static Analysis}
First, we examine the load balancing efficiency using the Gini coefficient, where 0 indicates perfect balance, and 1 represents maximum imbalance. We put the overall average Gini index in relation with the overall mean of the average QoS for each strategy, and we show in Figure~\ref{fig:scatter-sparse-avg_gini_qos} their interaction. Intuitively, the top-left strategies are the lowest performing ones as higher Gini index corresponds to worse load balancing efficiency and the QoS is lower; the bottom-right corner includes instead the better performing strategies.
In the full capacity scenario~(Fig.~\ref{fig:scatter-sparse-full-morning-avg_gini_qos}) we have basically perfect QoS for every strategy; however, we can notice how the benchmark ones show worse load balancing capabilities. Differently, when the RSU capacity is halved, as depicted in Fig.~\ref{fig:scatter-sparse-half-morning-avg_gini_qos}, the benchmarks start showing limitations also on the QoS. On the flip side, MACH strategies, especially with shorter load sharing intervals, performs better. Overall, the latest handover strategy is the one that leads to worst load balancing, potentially leading to bottlenecks in more crowded scenarios. Conversely, the ARHC-Oracle and ARHC-01s offer the best performance altogether.
Interestingly, load balancing is slightly worse with increased RSU capacity rather with half capacity. However, this difference is negligible.

\begin{figure}
    \begin{subfigure}[b]{0.43\linewidth}
    \centering
    \includegraphics[width=\linewidth]{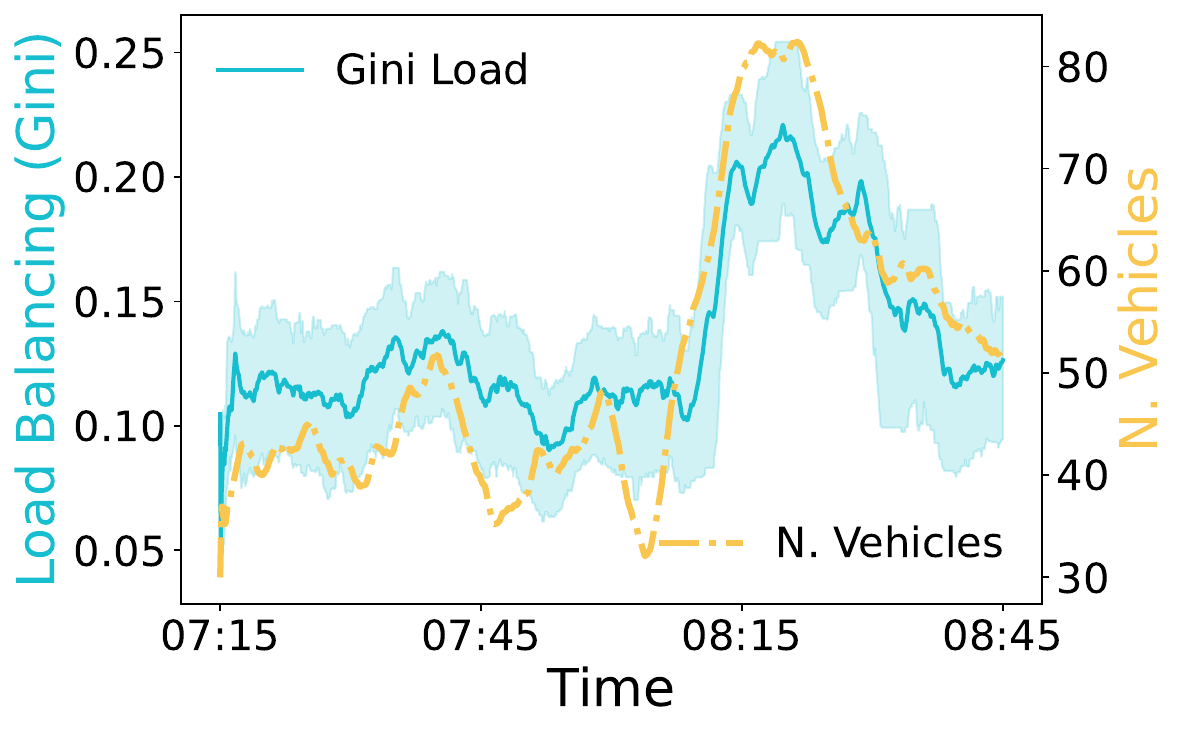}    
    \caption{Full capacity.}
    \end{subfigure}
    \hspace{5pt}
    \begin{subfigure}[b]{0.43\linewidth}
        \centering
    \includegraphics[width=\linewidth]{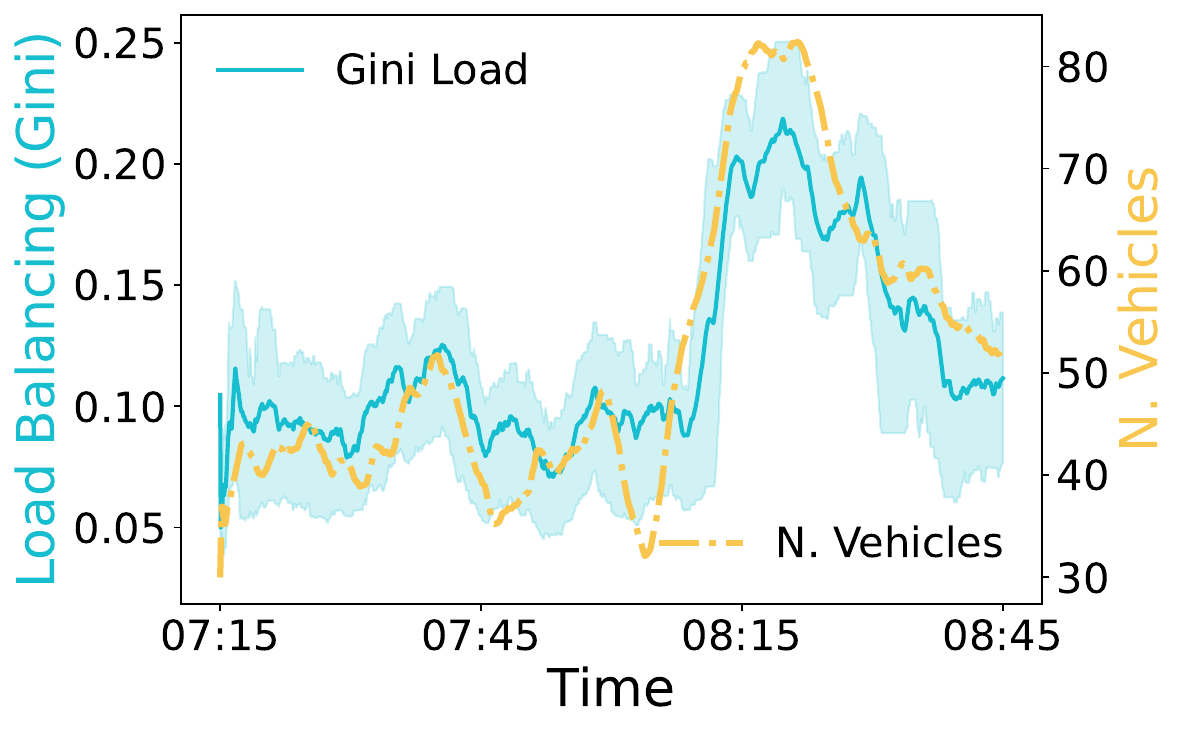}
    \caption{Half capacity.}
    \end{subfigure}
    \caption{25th and 75th quartile, plus mean Gini Index over time at full and half capacity in the sparse RSUs configuration.}
    \label{fig:gini-sparse-time}
\end{figure}

\begin{figure}
    \begin{subfigure}[b]{0.43\linewidth}
    \centering
    \includegraphics[width=\linewidth]{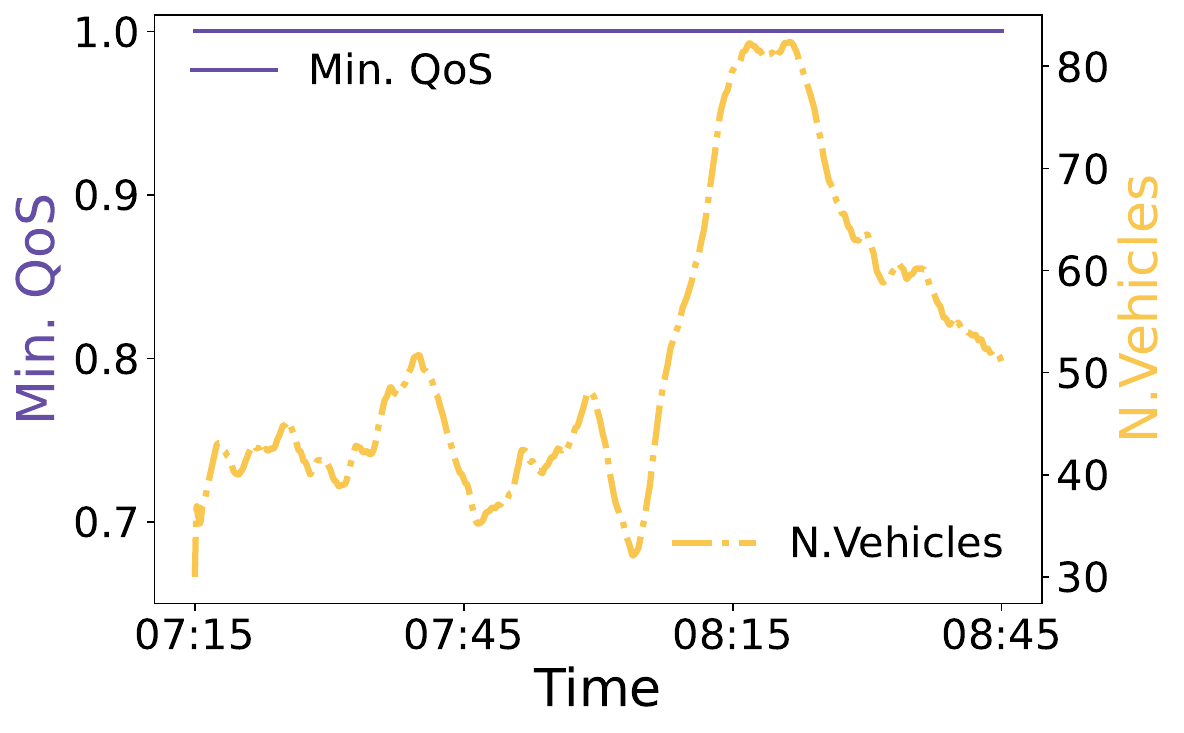}    
        \caption{Full capacity.}
            \label{fig:qos-sparse-time-full}
    \end{subfigure}
    \hspace{5pt}
    \begin{subfigure}[b]{0.43\linewidth}
        \centering
    \includegraphics[width=\linewidth]{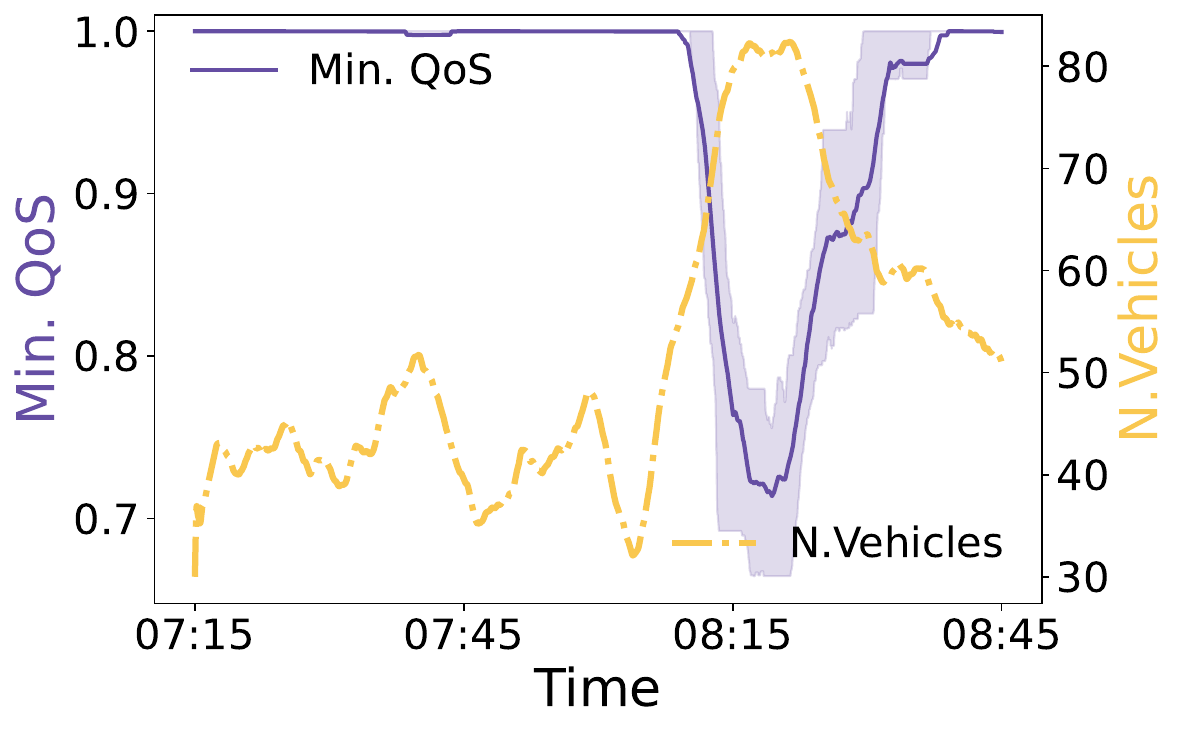}
        \caption{Half capacity.}
        \label{fig:qos-sparse-time-half}
    \end{subfigure}
    \caption{25th and 75th quartile, minimum QoS value over time at full and half capacity in the sparse RSUs configuration.}
    \label{fig:qos-sparse-time}
\end{figure}

\paragraph{Dynamic Analysis}
In addition, we explore how our performance metrics vary over time, and in relation to the dynamic traffic conditions. For the sake of conciseness, here we focus on the ARHC-01s strategy as it offers good performance and presents a reasonable RSUs update rate. Figure~\ref{fig:gini-sparse-time} depicts how the average Gini Index evolves. The progression clearly shows how an increase in the number of vehicles impacts the RSU capacity of load balancing.
Furthermore, we evaluate the impact of traffic volume in the QoS that the RSU can achieve. The results showed in Fig.~\ref{fig:qos-sparse-time} represent the minimum QoS values, confirming the intuition that heavier traffic is a key factor in QoS degradation, particularly under reduced RSU capacity, as shown in Fig.~\ref{fig:qos-sparse-time-half} (cfr. traffic density in Fig.~\ref{fig:6_eval:creteil_morning_density}).  These findings emphasize the importance of strategic RSU placement and capacity planning to mitigate the effects of congestion on service quality.

\subsection{Scenario 2: Créteil Roundabout, Dense Configuration} \label{sec:exp:creteil_dense}

In this section, we examine the performance of the \textit{dense} RSU configuration, evaluating how adding more nodes improves the infrastructure capacity to handle higher traffic volumes and more complex handover scenarios. 

\subsubsection{Handover Frequency}

\begin{figure}
    \centering
    \begin{subfigure}[b]{0.45\textwidth}
        \centering
        \includegraphics[width=\linewidth]{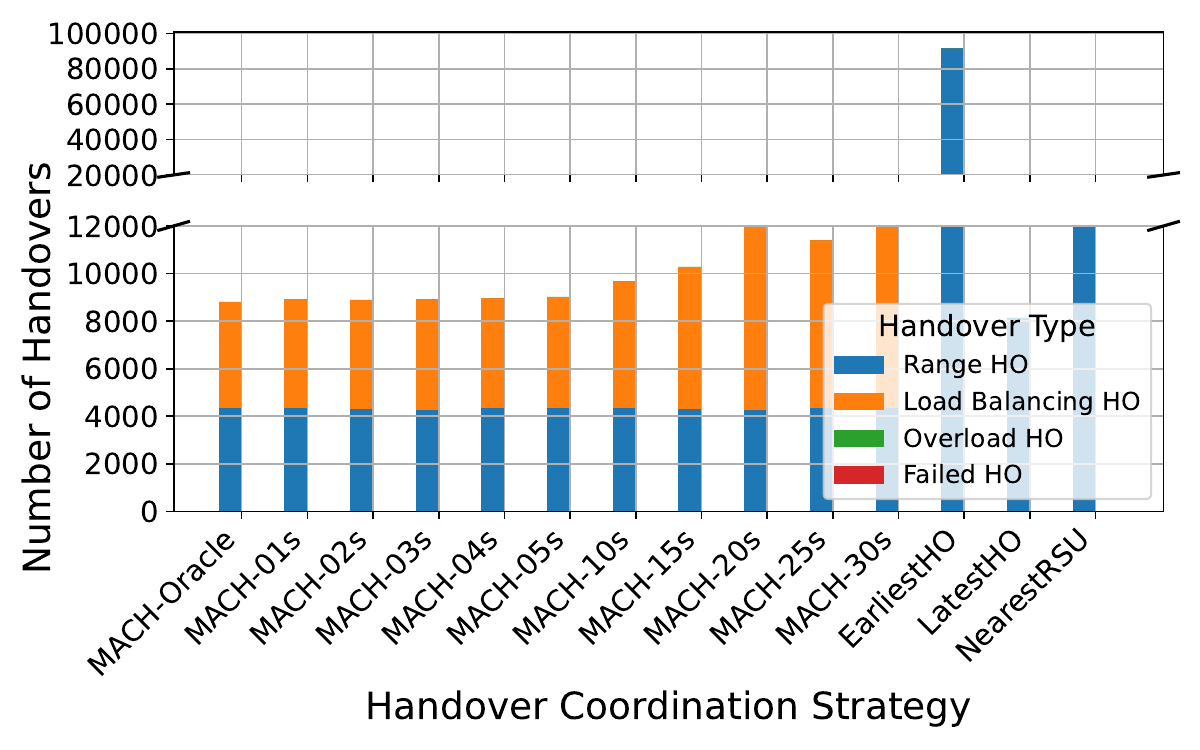}
        \caption{Full RSU Capacity}
        \label{fig:6_eval:res:creteil_morning_dense_full_ho}
    \end{subfigure}
    \hspace{5pt}
    \begin{subfigure}[b]{0.45\textwidth}
        \centering
        \includegraphics[width=\linewidth]{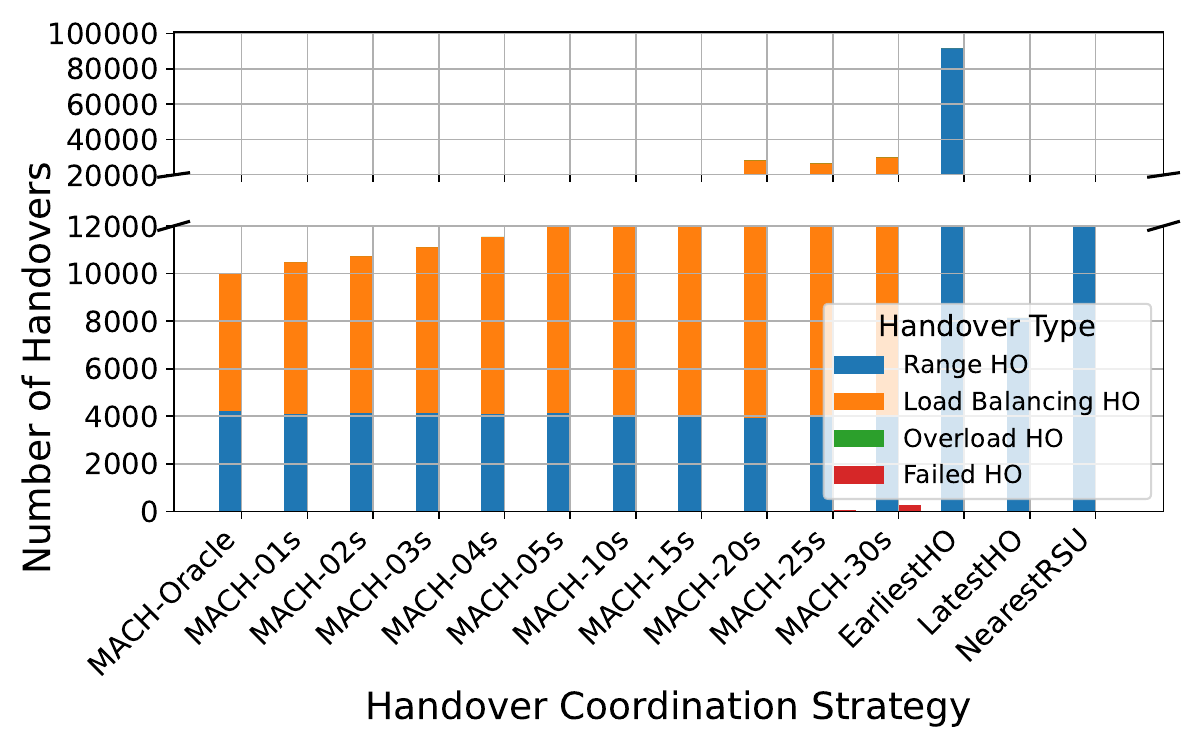}
        \caption{Half RSU Capacity}
        \label{fig:6_eval:res:creteil_morning_dense_half_ho}
    \end{subfigure}
    \hfill
    \caption{Number of Handovers of Créteil Dense Configuration at Morning}
    \label{fig:6_eval:res:creteil_morning_dense_ho}
\end{figure}


Figure~\ref{fig:6_eval:res:creteil_morning_dense_ho} illustrates the number of handovers across full, and half RSU capacities.
At first glance, we can notice how the earliest handover strategy requires the highest number of handovers, significantly more than any other strategy for both capacity configurations.
The latest handover strategy consistently minimizes handovers, outperforming the other strategies regardless of the mobility trace. MACH generally performs better than the nearest RSU strategy at full RSU capacity and remains competitive at low load-sharing intervals under half capacity. However, its performance deteriorates as load-sharing intervals increase, with most handovers attributed to load balancing in these cases. Furthermore, there are no overload or failed handovers in this scenario; a reasonable conclusion is that adding more nodes reduces the frequency of failures.


\subsubsection{Performance Metrics: QoS and Load Balancing Efficiency}



\begin{figure}[h]
    \begin{subfigure}[b]{0.43\linewidth}
        \centering
        \includegraphics[width=\linewidth]{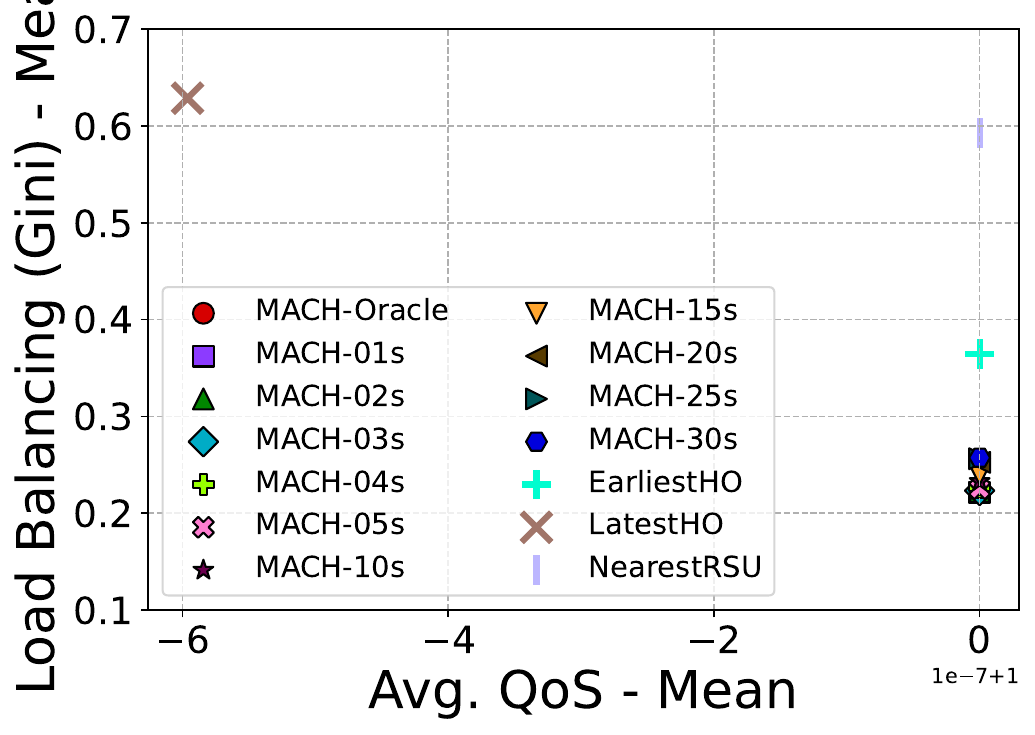}
        \caption{Full RSU Capacity.}
        \label{fig:scatter-dense-full-morning-avg_gini_qos}
    \end{subfigure} 
    \hspace{5pt}
    \begin{subfigure}[b]{0.43\linewidth}
        \centering
        \includegraphics[width=\linewidth]{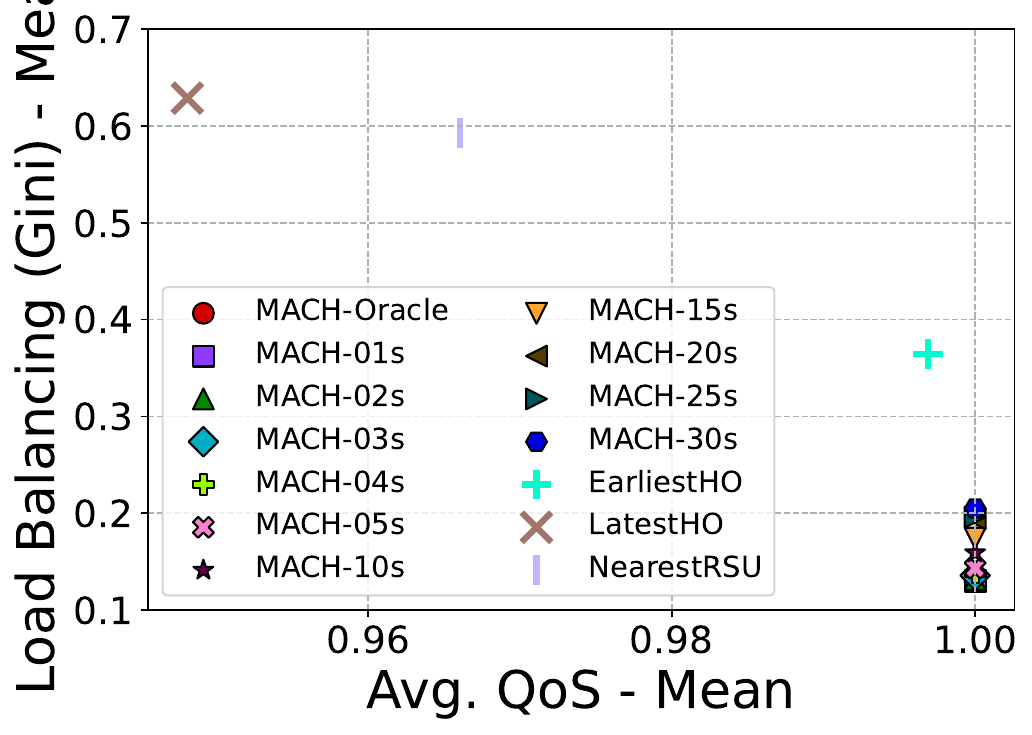}
        \caption{Half RSU Capacity.}
        \label{fig:scatter-dense-half-morning-avg_gini_qos}
    \end{subfigure}
    \caption{Scatter plot of load balancing and QoS for the dense RSU scenario.}
    \label{fig:scatter-dense-avg_gini_qos}
\end{figure}

\paragraph{Static Analysis}
The dense RSU configuration reflects trends similar to those of the sparse setup. At full RSU capacity, both MACH and baseline strategies achieve optimal values. In Fig.~\ref{fig:scatter-dense-full-morning-avg_gini_qos} we can notice how the variation is imperceptible if not for the zoom in. However, as RSU capacity decreases (Fig.~\ref{fig:scatter-dense-half-morning-avg_gini_qos},) MACH outperforms baseline strategies; still, the earliest handover still shows almost perfect average QoS values. For what concerns the load balancing efficiency, MACH strategies largely outperform the baselines, though the earliest handover strategy remains the best performer.. Still, similarly to the sparse configuration, MACH achieves better load distribution in more resource-constrained scenarios. Higher load-sharing intervals consistently lead to increased imbalance, with this effect being more pronounced in lower-capacity configurations. 

\paragraph{Dynamic Analysis}
\begin{figure}
    \begin{subfigure}[b]{0.43\linewidth}
    \centering
    \includegraphics[width=\linewidth]{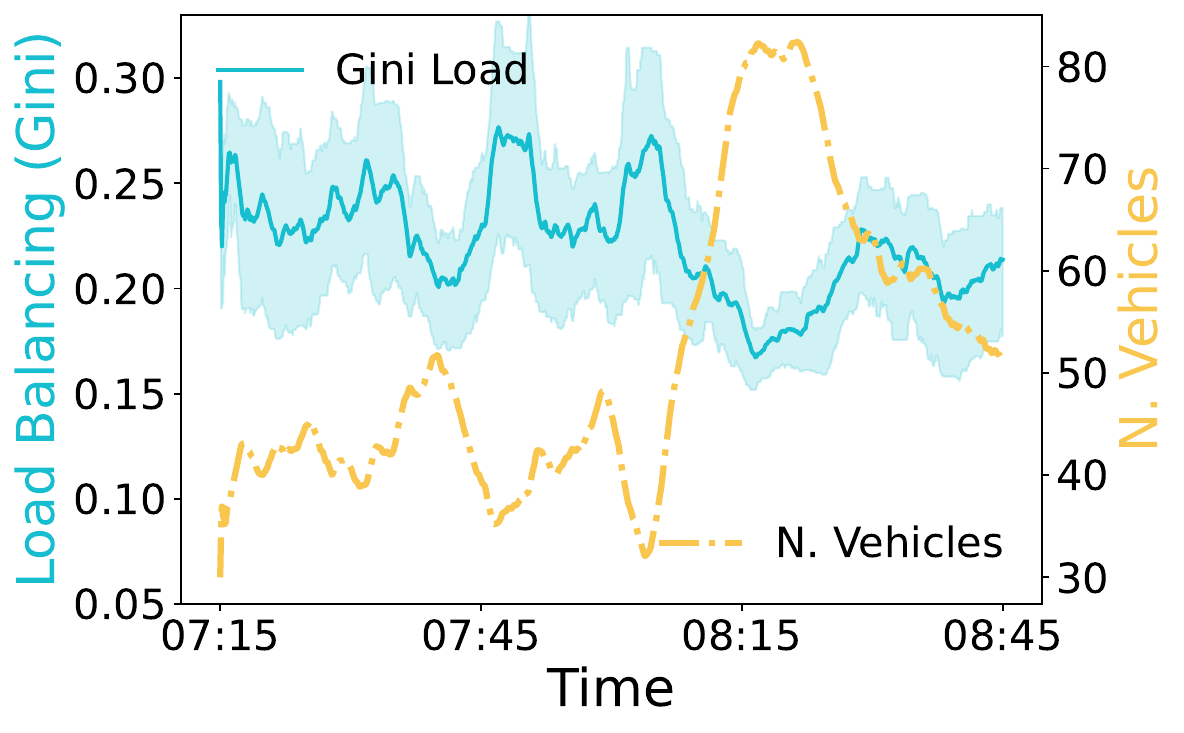}    
    \caption{Full capacity.}
    \label{fig:gini-dense-time-full}
    \end{subfigure}
    \hspace{5pt}
    \begin{subfigure}[b]{0.43\linewidth}
        \centering
    \includegraphics[width=\linewidth]{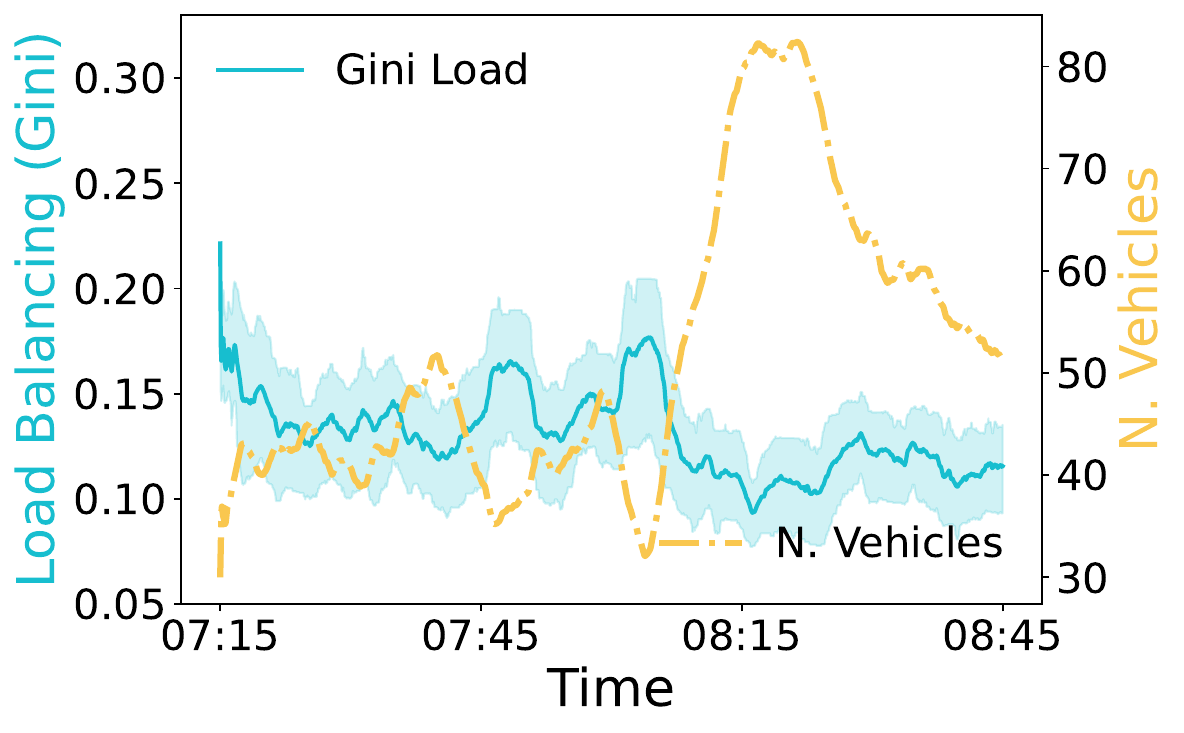}
    \caption{Half capacity.}
    \label{fig:gini-dense-time-half}
    \end{subfigure}
    \caption{25th and 75th quartile, plus mean Gini Index over time at full and half capacity in the dense RSUs configuration.}
    \label{fig:gini-dense-time}
\end{figure}

\begin{figure}
    \begin{subfigure}[b]{0.4\linewidth}
    \centering
    \includegraphics[width=\linewidth]{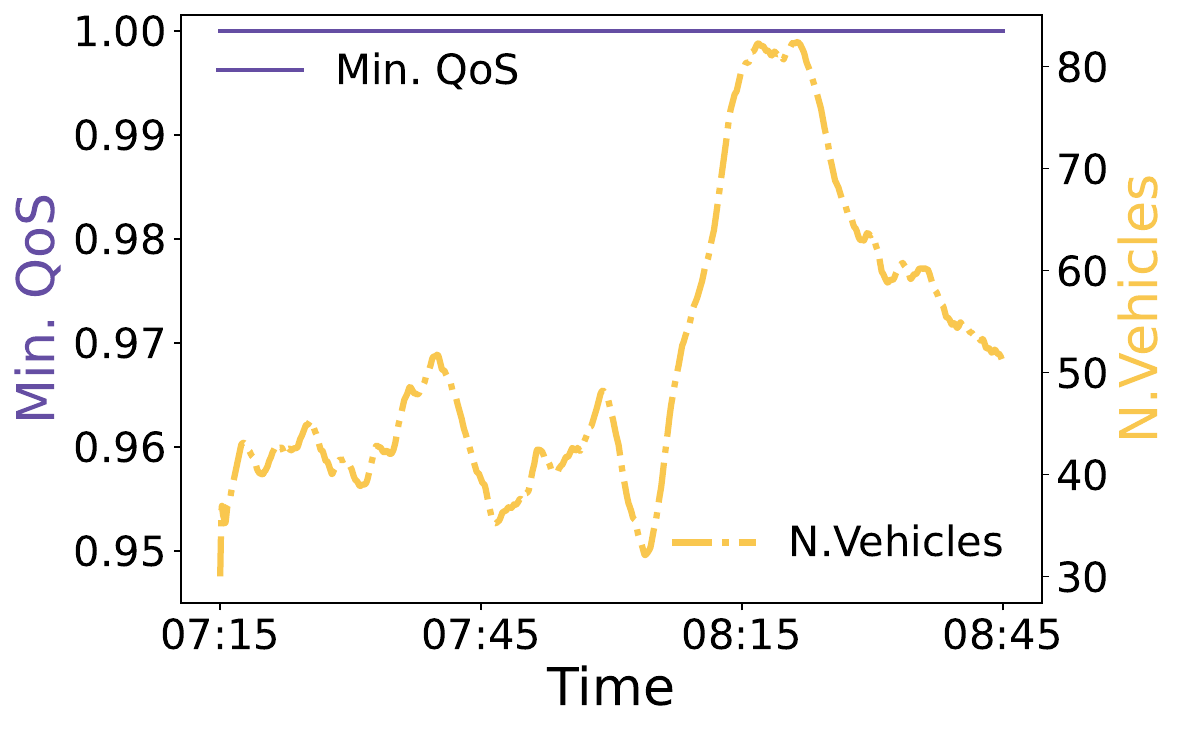}    
    \caption{Full capacity.}
    \label{fig:qos-dense-time-full}
    \end{subfigure}
    \hspace{5pt}
    \begin{subfigure}[b]{0.4\linewidth}
        \centering
    \includegraphics[width=\linewidth]{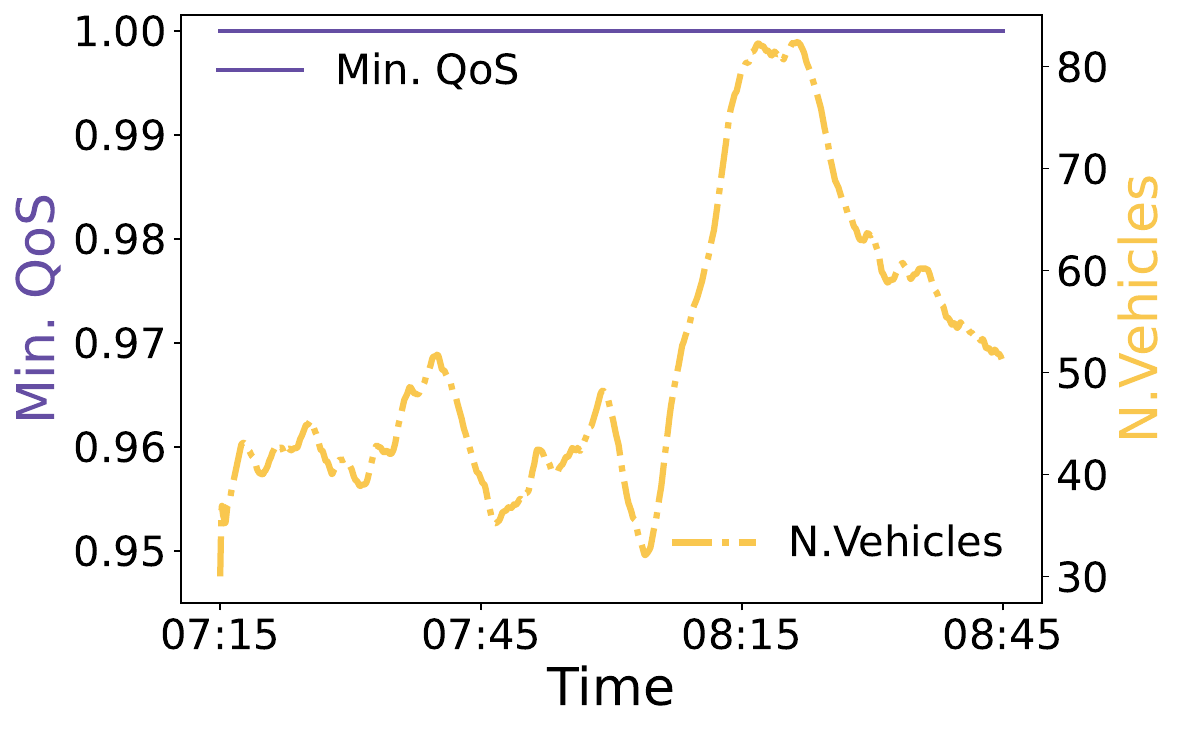}
    \caption{Half capacity.}
    \label{fig:qos-dense-time-half}
    \end{subfigure}
    \caption{25th and 75th quartile, minimum QoS value over time at full and half capacity in the dense RSUs configuration.}
    \label{fig:qos-dense-time}
\end{figure}

In the dense RSU configuration, the presence of 9 RSUs, with overlapping coverage areas, introduces additional complexity in managing load distribution and maintaining QoS. Still, the MACH strategies shows promising results both in full and half capacity of the RSUs. Still, as presented in Figure~\ref{fig:gini-dense-time} it encounters more significant challenges during periods of low traffic. For what concerns the QoS, both at full and half capacity the minimum values are still perfect, as shown in Fig.\ref{fig:qos-dense-time}. Overall, in the best-case scenario, the Gini coefficient rises during low traffic but remains manageable during peak periods, highlighting the strategy’s capability to balance load under heavier traffic.

\subsection{Scenario 3: Impact of RSU Failure on Strategy Performance}

This experiment examines the impact of an RSU failure on the performance of the MACH strategy and baseline approaches. The scenario uses the sparse 4-RSU setup from Experiment 1, with the south RSU (red) deliberately disabled to simulate a failure (see Figure~\ref{fig:6_eval:creteil_4_rsu_config}), i.e., the one that receives the highest load of traffic. This failure scenario is tested under full and half RSU capacity settings, analyzing the morning traffic trace.


\subsubsection{Handover Frequency}

\begin{figure}
    \centering
    \begin{subfigure}[b]{0.45\textwidth}
        \centering
        \includegraphics[width=\linewidth]{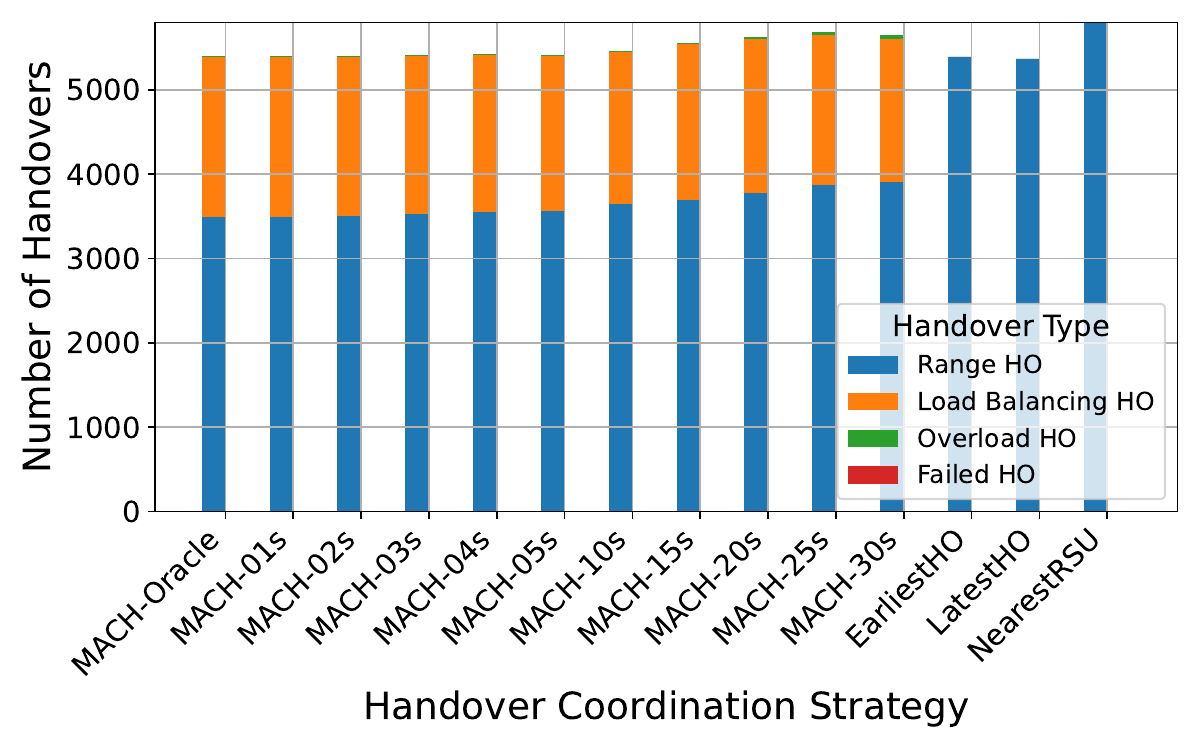}
        \caption{Full RSU Capacity}
        \label{fig:6_eval:res:exp3_morning_full_ho}
    \end{subfigure}
    \hspace{5pt}
    \begin{subfigure}[b]{0.45\textwidth}
        \centering
        \includegraphics[width=\linewidth]{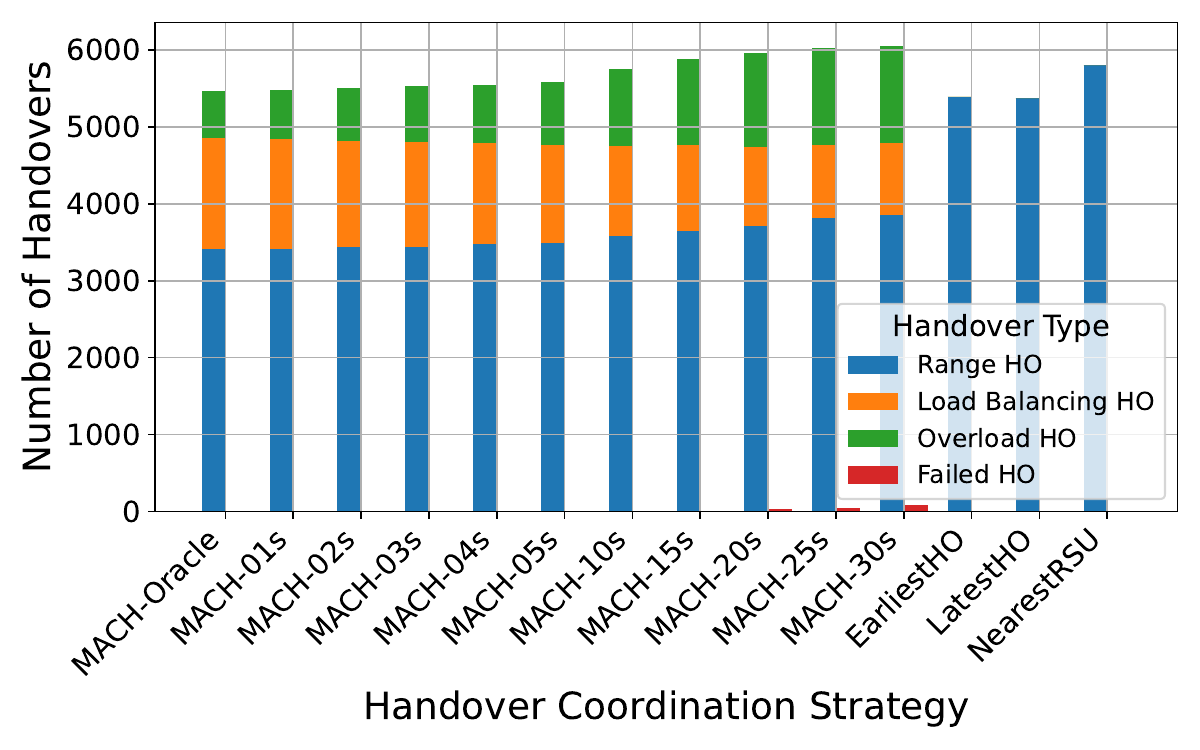}
        \caption{Half RSU Capacity}
        \label{fig:6_eval:res:exp3_morning_half_ho}
    \end{subfigure}
    \caption{Number of Handovers of Créteil Failure Configuration at Morning}
    \label{fig:6_eval:res:exp3_morning_half_ho_full}
\end{figure}

Figure~\ref{fig:6_eval:res:exp3_morning_half_ho_full} shows handover counts during the morning trace for both full and half RSU capacities in the failure scenario.
Handovers are relatively consistent across configurations at full capacity, with around 5500 total handovers. Most handovers are triggered by vehicles moving out of range in the MACH setup, showing efficient management as vehicles pass through RSUs coverage areas. No failed handovers occur, reflecting effective load distribution despite the RSU failure.
In the half-capacity scenario, the pattern remains similar, but overload-triggered handovers appear, showing the system’s efforts to balance load among the remaining RSUs. Failed handovers are minimal, even when the load-sharing intervals are high.
These findings demonstrate the MACH strategy’s resilience in managing handovers effectively, even with a disabled RSU. However, due to the road layout, the low RSU density limits handover options, leading to a strict order of handovers. Increasing RSU density could improve flexibility.
One limitation is that MACH only considers handovers to RSUs within a vehicle’s range. If a vehicle exits a station's range and loses coverage, it won't reconnect until it enters another station’s range, even if a closer RSU exists. Addressing this could improve the system’s ability to handle unexpected disruptions.
Overall, the experiment highlights the MACH strategy's ability to maintain operational integrity under challenging conditions, with opportunities for future enhancements in adaptability and coverage management.

\subsubsection{Performance Metrics: QoS and Load Balancing Efficiency}


\begin{figure}[h]
    \begin{subfigure}[b]{0.43\linewidth}
        \centering
        \includegraphics[width=\linewidth]{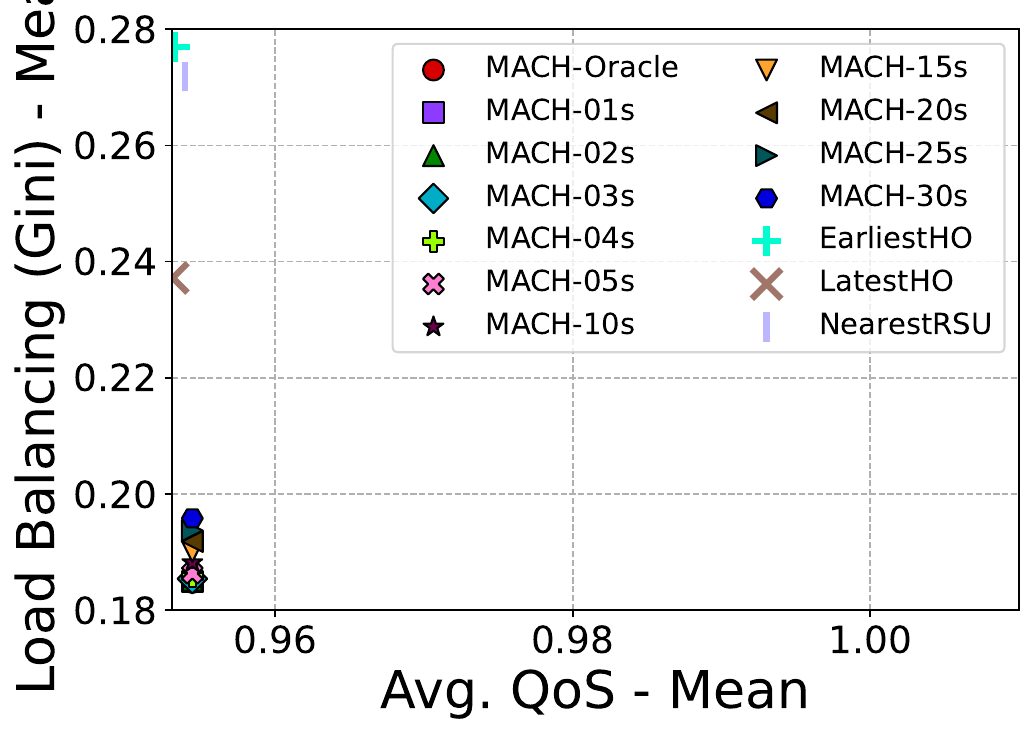}
        \caption{Full RSU Capacity.}
        \label{fig:scatter-faiil-full-morning-avg_gini_qos}
    \end{subfigure} 
    \hspace{5pt}
    \begin{subfigure}[b]{0.43\linewidth}
        \centering
        \includegraphics[width=\linewidth]{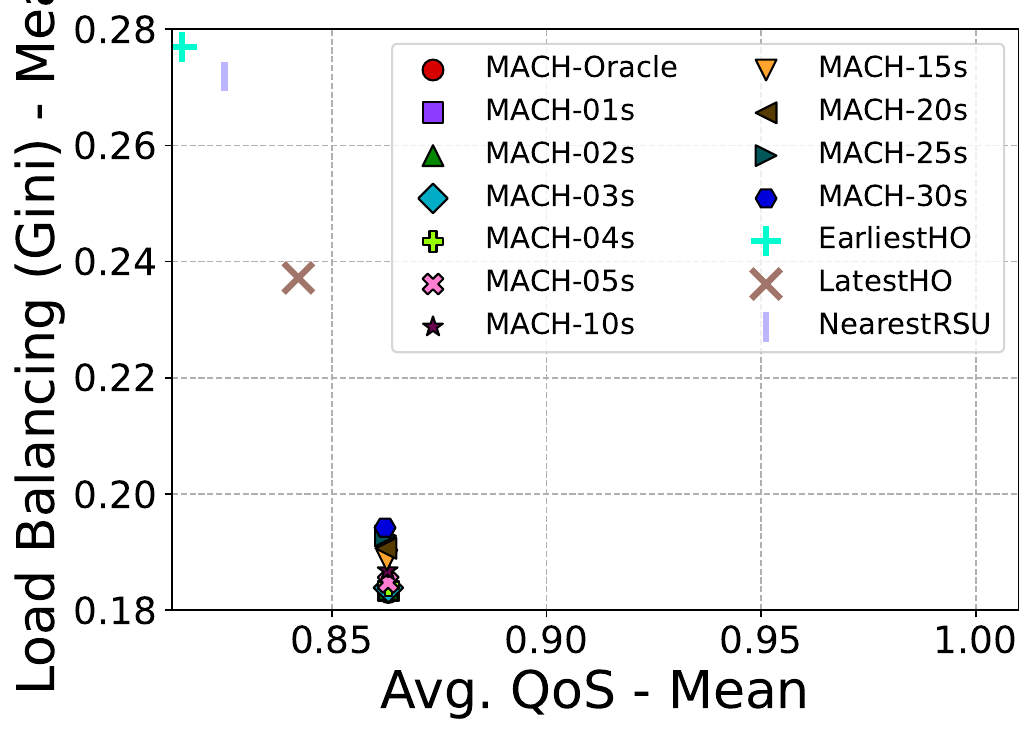}
        \caption{Half RSU Capacity.}
        \label{fig:scatter-fail-half-morning-avg_gini_qos}
    \end{subfigure}
    \caption{Scatter plot of load balancing and QoS for the RSU failure scenario.}
    \label{fig:scatter-fail-avg_gini_qos}
\end{figure}

\paragraph{Static Analysis}
The RSU failure mostly impacts the QoS, as shown in Fig.~\ref{fig:scatter-fail-avg_gini_qos}, revealing important insights into the resilience of the MACH strategy compared to baseline approaches. Under full capacity, the network maintained a relatively high average QoS. This suggests the system was robust enough to sustain decent overall service quality. The similar performance of the MACH strategy and baseline approaches under full capacity conditions suggests that the high amount of available RSU capacity reduced the need for advanced load balancing, leading to comparable results.
In contrast, when the RSUs operated at half capacity, as shown in Fig.~\ref{fig:scatter-fail-half-morning-avg_gini_qos}, the MACH strategy outperformed the baseline approaches, showing its effectiveness in balancing the load and preserving QoS as much as possible under constrained conditions. As vehicles exited the coverage area of the overloaded western (blue) RSU (see Figure~\ref{fig:6_eval:creteil_4_rsu_config}), they often could not connect to another station if none was within immediate range, leading to localized service degradation. This issue was particularly severe for vehicles that exited the roundabout using the southern direction or entered from the south, where the lack of an alternative connection point further increased the load on the remaining RSUs.
Regarding the load balancy efficiency, the MACH strategy outperformed the baseline approaches, maintaining a more balanced load distribution. Among the baseline strategies, the latest handover (handover) strategy performed the best but still fell short compared to the MACH strategy's effectiveness in managing load imbalances
Overall, while the MACH strategy showed strong performance under full capacity, its limitations became more apparent in the half-capacity scenario, especially in dealing with RSU failures. Future improvements could enhance the handover logic to consider potential connections to nearby but out-of-range stations, increasing the system's resilience in failure scenarios.

\begin{figure}
    \begin{subfigure}[b]{0.43\linewidth}
    \centering
    \includegraphics[width=\linewidth]{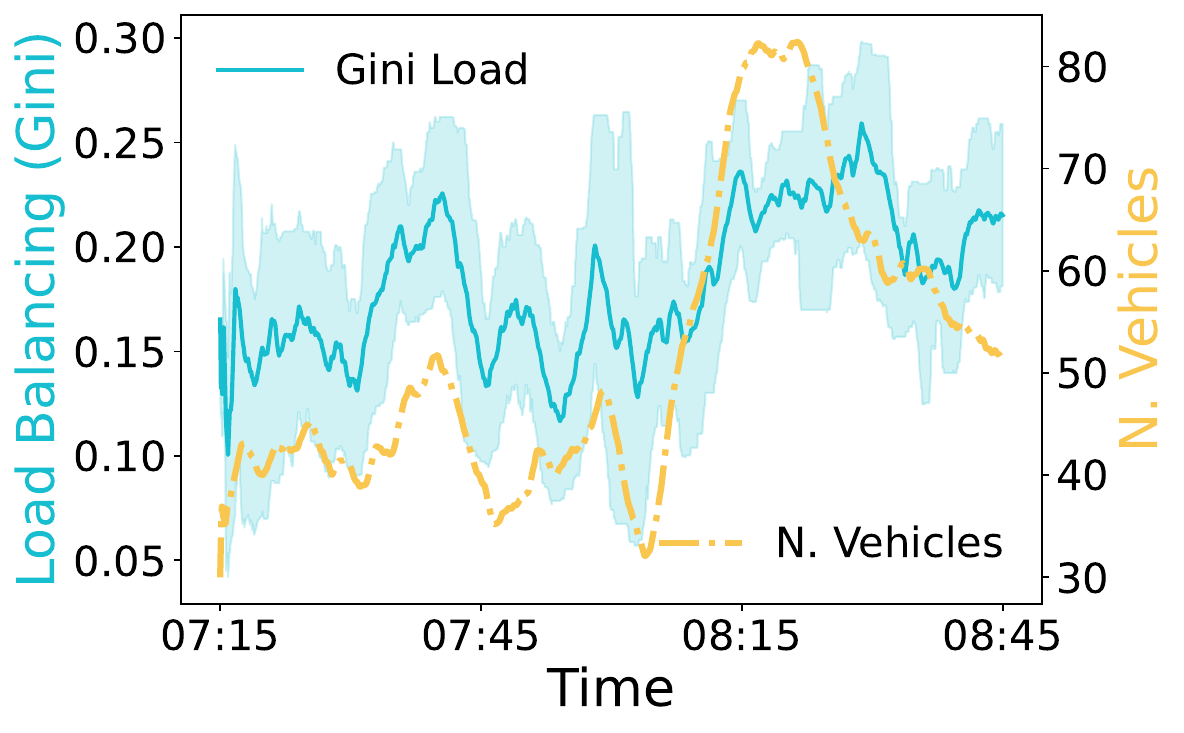}    
    \caption{Full capacity.}
    \label{fig:gini-fail-time-full}
    \end{subfigure}
    \hspace{5pt}
    \begin{subfigure}[b]{0.43\linewidth}
        \centering
    \includegraphics[width=\linewidth]{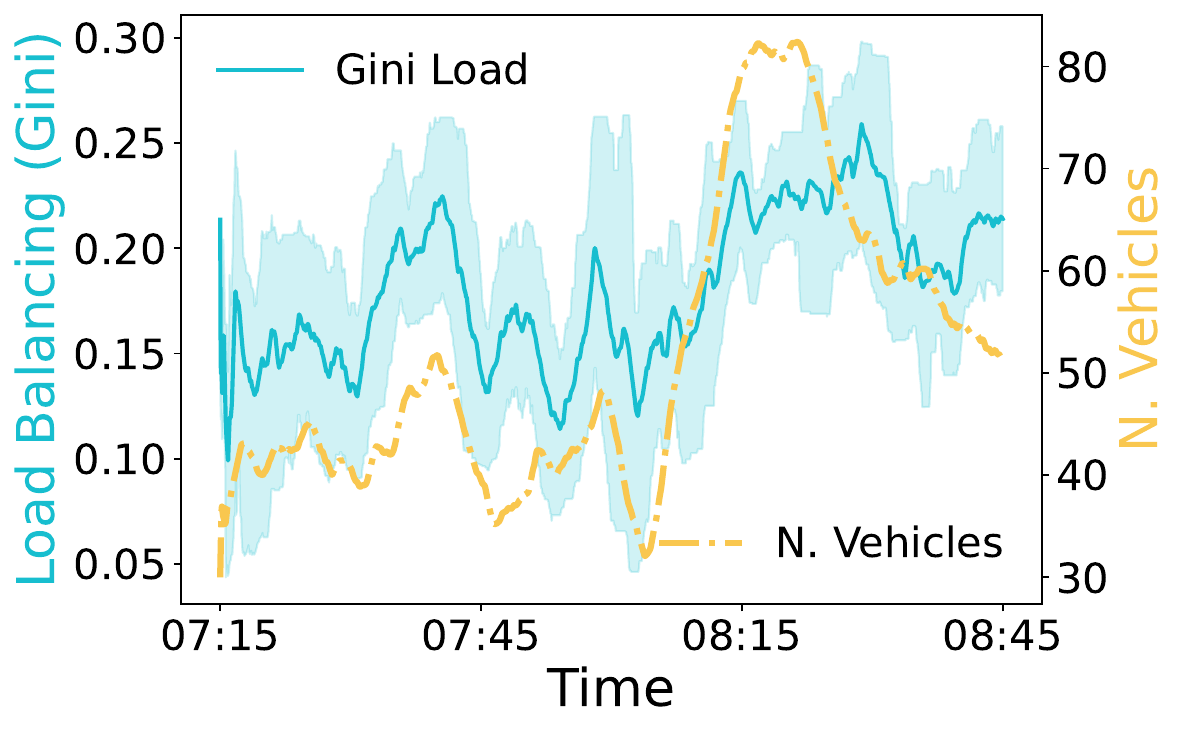}
    \caption{Half capacity.}
    \label{fig:gini-fail-time-half}
    \end{subfigure}
    \caption{25th and 75th quartile, plus mean Gini Index over time at full and half capacity in the dense RSUs configuration.}
    \label{fig:gini-fail-time}
\end{figure}

\begin{figure}
    \begin{subfigure}[b]{0.43\linewidth}
    \centering
    \includegraphics[width=\linewidth]{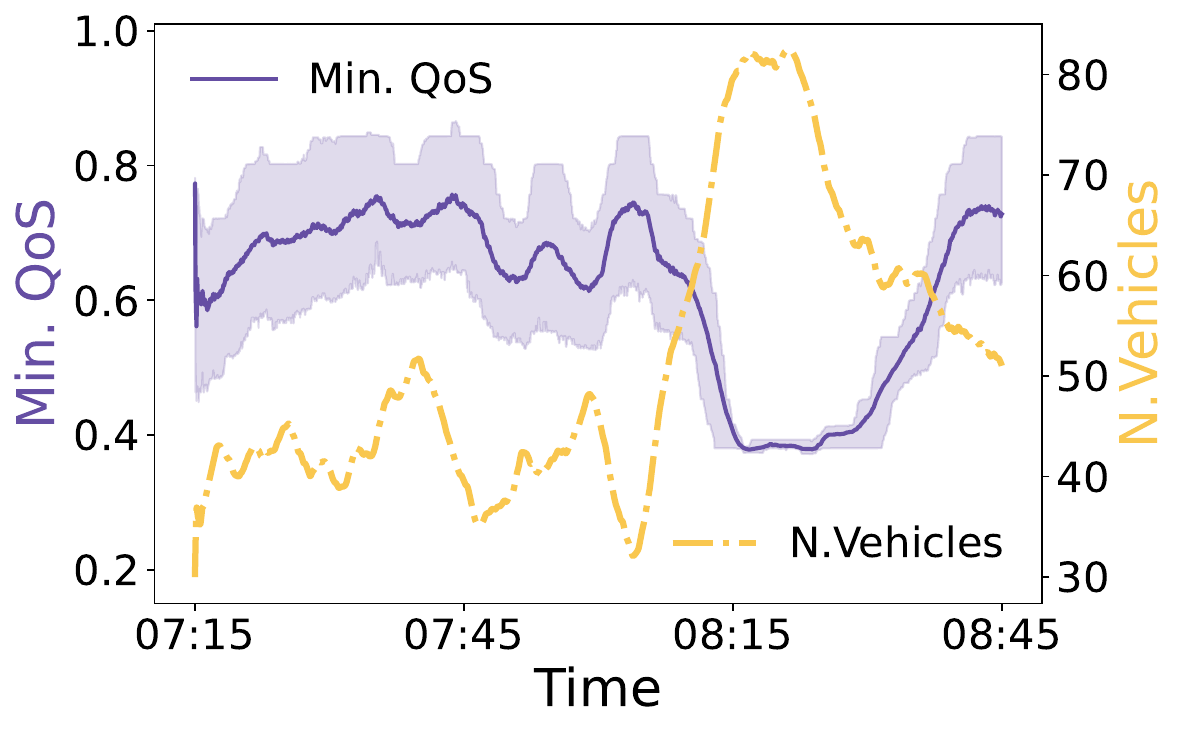}    
    \caption{Full capacity.}
    \label{fig:qos-fail-time-full}
    \end{subfigure}
    \hspace{5pt}
    \begin{subfigure}[b]{0.43\linewidth}
        \centering
    \includegraphics[width=\linewidth]{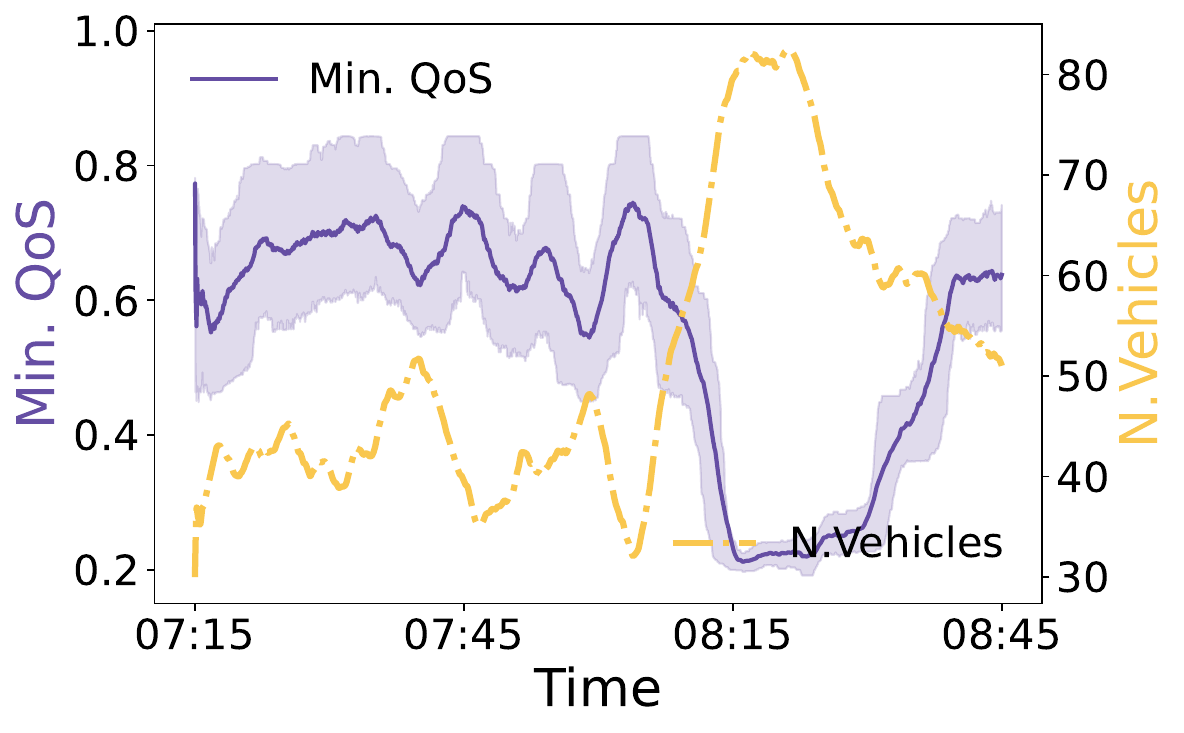}
    \caption{Half capacity.}
    \label{fig:qos-fail-time-half}
    \end{subfigure}
    \caption{25th and 75th quartile, minimum QoS value over time at full and half capacity in the dense RSUs configuration.}
    \label{fig:qos-fail-time}
\end{figure}

\paragraph{Dynamic Analysis}
At full RSU capacity, the Gini coefficient (Fig.\ref{fig:gini-fail-time-full}) ranges mostly between 0.1 and 0.25, with occasional peaks reaching up to 0.37. Although higher traffic correlates with some increases in load imbalance, these variations are not significantly more pronounced during peak traffic compared to other times. These values are similar in the half capacity scenario, as depicted in Fig.\ref{fig:gini-fail-time-half}, primarily due to the limited options for executing effective handovers.
The QoS metrics present a more critical picture. Looking at the minimum QoS over time (Figure \ref{fig:qos-fail-time-full}), its values are stable for much of the time but drop significantly during periods of high traffic. When the number of vehicles exceeds approximately 60, the minimum QoS decreases to around 30\%. More insights, which we could not fit in the paper, show that these declines are mainly driven by distance-based QoS issues, as vehicles move out of the coverage range of any RSU. In the half capacity scenario, as visualized in Fig.~\ref{fig:qos-fail-time-half}, while minimum QoS is relatively stable during low traffic, it drops sharply to around 20\% during peak traffic periods. 
This suggests a general system overload during these times, with some RSUs operating beyond their capacity and no other RSUs available to distribute the load effectively. Overall, while load distribution inequality remains consistent due to the limited handover options, QoS is much more affected at half capacity. When operating with reduced capacity, the system struggles to maintain satisfactory QoS levels during peak traffic, mainly because of the lack of available RSUs to manage the increased load effectively. This analysis underscores the critical importance of maintaining adequate RSU capacity to ensure stable QoS, particularly during periods of high traffic.

\section{Discussion and Broader Implications}
\label{sec:discussion}
This section synthesizes the key findings from the sparse and dense scenarios, highlighting the strengths and limitations of MACH under varying network conditions.
\subsection{Insights from the Evaluation Scenarios}
\paragraph{Effectiveness of MACH in Managing Handovers} 
In sparse and dense configurations, MACH demonstrates a strong ability to manage handovers efficiently, particularly under full-capacity conditions. The strategy’s decentralized approach, leveraging RSU-based decision-making, enables it to control handover processes, minimizing unnecessary disruptions and maintaining network stability. However, as RSU capacity decreases, the increased complexity in dense configurations introduces more frequent handovers, driven primarily by the need for aggressive load balancing. This was particularly evident during periods of high traffic, where the rise in vehicle numbers required more frequent and critical handovers in the sparse setup to maintain service continuity. This indicates that while MACH is highly effective in environments with limited infrastructure (such as the sparse configuration), its application in denser networks requires careful management to prevent excessive handover frequency, which could otherwise lead to increased network overhead and reduced QoS.

\paragraph{QoS Maintenance and Load Balancing Efficiency} 
The performance of MACH in maintaining QoS is significantly influenced by the total network capacity and the distribution of RSUs. The total capacity is inherently lower in the sparse configuration due to the smaller number of RSUs (4 compared to 9 in the dense configuration). This limitation becomes more pronounced as RSU capacity is reduced, making it difficult for MACH to maintain high QoS levels, especially in high-traffic scenarios.
Interestingly, load balancing tends to improve in reduced-capacity scenarios, as indicated by the lower Gini index. This happens because all RSUs, having the same capacity, reach similar load levels when fully utilized. The MACH strategy responds by balancing the load more frequently to prevent overloads, leading to a more even distribution of the load across the network. This behavior highlights a key relationship between RSU capacity and load distribution: a reduced capacity can lead to better load balance, though it may come at the cost of overall QoS if the system's total capacity is insufficient.
In contrast, the dense configuration benefits from a higher total capacity and more overlapping RSU coverage, which enhances the system’s ability to manage peak loads and maintain higher QoS. 


\paragraph{Generalizability and Broader Applicability}
While MACH has been validated in the context of a single urban roundabout, the core principles of the system—decentralized negotiation, local utility estimation, and agent-based coordination—are designed to generalize across different vehicular mobility topologies. The roundabout scenario was chosen as a stress test environment with high handover density and dynamic service availability. However, the MACH protocol and agent interactions do not rely on topology-specific assumptions and are applicable to larger urban grids, highway RSU deployments, or hybrid environments combining mobile and static edge nodes. Future work includes validating MACH across a broader set of real-world maps and vehicle traces.

\paragraph{Limitations of the Evaluation Approach}
While the results demonstrate the efficacy of MACH in various scenarios, several limitations must be acknowledged. Firstly, the \textbf{scope} of the simulation is confined to a relatively \textit{small urban area}. While this setting provides valuable insights, it may not fully capture the challenges encountered in more extensive or diverse vehicular networks, such as highways or rural areas. This limitation suggests that further research is necessary to assess MACH’s performance across various environments.
Secondly, the simulation assumes \textbf{consistent RSU availability} and \textbf{reliable communication}, overlooking potential real-world issues such as RSU failures or data transmission errors. The findings from the failure scenario highlight MACH’s dependence on RSU availability, where the loss of an RSU led to significant QoS degradation, particularly in sparse configurations where redundancy is minimal. This reliance on uninterrupted RSU operation underscores the importance of enhancing the strategy’s robustness to handle such disruptions.
Thirdly, the simplified \textbf{load-offloading model} used in the simulations focuses \textit{solely on computational load}, measured in G\gls{flops}, without accounting for other factors like memory usage or specific task requirements. 
MACH's coordination logic is intentionally lightweight and heuristic-based to support real-time decision-making at the RSU level. While this favors deployability and responsiveness, it does not capture long-term global optimality or provable performance bounds. 
In future work, we plan to integrate formal analytical models, such as Markov Decision Processes or constrained optimization frameworks, to complement the system-level design. Additionally, exploring MACH's behavior under larger-scale deployments with heterogeneous RSU capabilities and workload patterns remains an open area for research.

\paragraph{Takeaways} 
Overall, the performance of MACH across sparse and dense RSU configurations reveals that the strategy is robust and effective. Still, its success is highly dependent on the network’s total capacity and the distribution of RSUs. With its higher capacity and overlapping RSU coverage, the dense configuration provides a more favorable environment for MACH, enabling better management of peak loads and higher overall QoS. In contrast, the sparse configuration’s limited capacity presents challenges that require careful management to avoid significant QoS degradation. These findings underscore the importance of optimizing network topology and RSU distribution to maximize the effectiveness of MACH in diverse VEC environments.

\subsection{Strengths and Shortcomings of the MACH Strategy}

One of the most significant strengths of the MACH strategy is its \textbf{scalability} across different network environments. The approach has proven effective in both sparse and dense RSU configurations, demonstrating its ability to adapt to varying network density and capacity levels. This scalability ensures that MACH can be applied in diverse VEC scenarios, ranging from small urban areas with limited RSUs to larger, more complex networks with high RSU density.
Secondly, he MACH strategy consistently maintains \textbf{high QoS}, particularly under full RSU capacity conditions. Even in scenarios where RSU capacity is reduced, MACH outperforms baseline strategies by effectively managing handovers and balancing computational loads, highlighting its robustness.
However, when faced with an RSU failure, the strategy's effectiveness is significantly compromised. The reduction in available network infrastructure led to noticeable QoS degradation, particularly in areas with sparse RSU coverage. This suggests the need for further enhancements to better handle such scenarios.

Furthermore, the Gini coefficient analysis across multiple experiments demonstrates that MACH achieves a more \textbf{balanced load distribution} among RSUs, especially in resource-constrained scenarios. By preventing RSU overloads and ensuring that no RSU becomes a bottleneck, MACH enhances the overall stability and efficiency of the network. This balance is crucial in maintaining consistent service levels for all vehicles within the network, particularly during periods of high traffic.
In the event of network disruptions, such as an RSU failure, maintaining this balance becomes even more critical. The absence of an RSU exacerbates load imbalances, particularly when the strategy does not account for handovers to out-of-range stations, which could mitigate these effects.

Finally, MACH performs well in coordinating handovers, \textbf{minimizing unnecessary handovers}, and preventing failures even in scenarios with reduced RSU capacity. This robust coordination is essential for maintaining uninterrupted service as vehicles move through different RSU coverage areas, ensuring that computational tasks are seamlessly offloaded without compromising QoS.
Nonetheless, in scenarios where the network infrastructure is compromised, such as during an RSU failure, the current handover logic of MACH could be improved by considering connections to nearby but out-of-range stations. This could help sustain network stability and service quality even under challenging conditions.

As a limitation, the effectiveness of MACH highly depends on the available RSU capacity. In scenarios where RSU capacity is significantly reduced, MACH's ability to maintain high QoS and balanced load distribution diminishes. Rapid increases in traffic could quickly lead to QoS drops due to limited RSUs, highlighting the strategy’s dependence on sufficient capacity; this limitation was particularly evident when the network faced an RSU failure.

\paragraph{Takeaways}
The MACH strategy demonstrates considerable strengths, including scalability, robust handover coordination, and effective QoS maintenance, making it a promising solution for VEC. However, the shortcomings identified in both normal operations and failure scenarios highlight areas where further research and development are needed. Addressing these limitations, particularly in expanding the simulation scope, optimizing MACH for low-capacity scenarios, and incorporating edge cases and failure scenarios, will enhance the strategy's applicability and robustness.

\section{Conclusion}
\label{sec:conclusion}
In this paper, we presented a multi-agent-based methodology for task offloading  between collaborative Road Side Units (RSUs) in a vehicular Edge computing scenario. 
%
To handle dynamic mobility patterns of vehicles and the complexity during real-time handovers, we presented a solution that allows multiple RSUs to make decentralized offloading decisions. To make informed decisions, RSUs continuously exchange information on their current load and consider the predicted trajectories of their connected vehicles. According to this, RSUs themselves initiate a handover if they expect that one of its clients could receive a better service from another RSU.
%
This represents a paradigm shift in handover coordination because decisions are made at the network edge with minimum latency. Considering that request patterns of vehicles and their generated load at RSUs varies greatly during the day, our approach can significantly enhance network adaptability and resilience.
To evaluate our ideas, we developed an accurate simulation environment based on real-world traffic data; as a result, our proposed methodology clearly achieved the highest quality of service, while finding the fairest load distribution between RSUs and ensuring minimum numbers of total handovers.
To further enhance the robustness of our approach, we plan introducing more stress to the simulation environment, such as as dynamic RSU failures or traffic bursts. Furthermore, we plan to investigate advanced algorithms for proactive handovers and resource allocation, embedding predictive decision-making, e.g., by having models predict network load based on historical data.

\begin{acks}
This work is funded by the HORIZON Research and Innovation Action 101135576 INTEND ``Intent-based data operation in the computing continuum.''
\end{acks}

\bibliographystyle{ACM-Reference-Format}
\bibliography{references,Boris}
\end{document}